\pdfoutput=1
\documentclass[amsmath, amsymb, 
aps,prc,reprint,groupedaddress,nofootinbib,longbibliography,abbrv]{revtex4-1}

\usepackage{graphicx}   % need for figures
\usepackage{color}
\usepackage[colorlinks=true,linkcolor=blue,citecolor=blue, urlcolor=blue]{hyperref}   % use for hypertext links, including those to external documents and URLs
\usepackage{bm} % for bold caligraphic font

\usepackage[raggedright,tight,nooneline]{subfigure}

\def\p{{\bf p}}

\def\st{\begin{equation}}
\def\stp{\end{equation}}
\def\bg{\begin{eqnarray}}
\def\nd{\end{eqnarray}}
\def\Eq#1{Eq.~(\ref{#1})}
\def\Eqs#1{Eqs.~(\ref{#1})}
\def\eq#1{(\ref{#1})}

\def\Fig#1{Fig.~\ref{#1}}
\def\Figs#1{Figs.~\ref{#1}}
\def\Sect#1{Sect.~\ref{#1}}
\def\snn{{\sqrt{s_{\scriptscriptstyle NN}}}}
\def\llangle{\left\langle}
\def\rrangle{\right\rangle}

\begin{document}

\title{Fluctuations of harmonic and radial flow in heavy ion collisions with principal components}

\author{Aleksas Mazeliauskas}
\email[]{aleksas.mazeliauskas@stonybrook.edu}
\affiliation{Department of Physics and Astronomy, Stony Brook University, Stony Brook, New York 11794, USA}
\author{Derek Teaney}
\email[]{derek.teaney@stonybrook.edu}
\affiliation{Department of Physics and Astronomy, Stony Brook University, Stony Brook, New York 11794, USA}

\date{\today}
%%%%%%%%%%%%%%%%%%%%%%%%%%%%%%%%% abstract.tex %%%%%%%%%%%%%%%%%%%%%%%%%%%%%%%%%

\begin{abstract}
    We analyze the spectrum of harmonic flow, $v_n(p_T)$ for $n=0\text{--}5$,
in event-by-event hydrodynamic simulations of Pb+Pb collisions 
at the CERN Large Hadron Collider ($\snn=2.76\,{\rm TeV}$) with principal 
component analysis (PCA). 
The PCA procedure finds two dominant contributions to the two-particle
correlation function. The leading component is identified with the event plane 
$v_n(p_T)$, while the subleading component is responsible for factorization 
breaking in hydrodynamics.   
For $v_0$, $v_1$, and $v_3$ the
subleading flow is a response to the radial 
excitation of the corresponding eccentricity. By contrast, for $v_2$ the subleading flow
in \emph{peripheral collisions} is dominated by the nonlinear mixing between the leading elliptic flow and radial flow fluctuations. 
In the $v_2$ case, the sub-sub-leading mode more closely reflects the response 
to the radial excitation of $\varepsilon_2$. 
A consequence of this picture is that the elliptic flow fluctuations
and factorization breaking change rapidly 
with centrality, and in central collisions (where the leading $v_2$ is 
small and nonlinear effects can be neglected) the subsub-leading mode
becomes important.
Radial flow fluctuations and nonlinear mixing also play a significant role in 
the factorization breaking of $v_4$ and $v_5$.
We construct good geometric predictors for the 
orientation and magnitudes of the leading and subleading flows 
based on a linear response to the geometry, and 
a quadratic mixing between the leading principal components. 
Finally, we suggest a set of measurements involving three point correlations
which can experimentally corroborate the nonlinear mixing of radial and 
elliptic flow and its important contribution to factorization breaking as a 
function of centrality.
\end{abstract}

\pacs{}

\maketitle

%%%%%%%%%%%%%%%%%%%%%%%%%%%%%%% introduction.tex %%%%%%%%%%%%%%%%%%%%%%%%%%%%%%%

\section{Introduction}

Two-particle correlation measurements are of paramount importance in studying
ultrarelativistic heavy ion collisions, and provide an extraordinarily
stringent test for theoretical models. Indeed, the measured two-particle
correlations exhibit elliptic, triangular, and higher harmonics flows,  which 
can be used to constrain the transport properties of the quark gluon plasma 
(QGP)~\cite{Heinz:2013th,Luzum:2013yya}.  The remarkable
precision of the experimental data 
as a function of transverse momentum and pseudorapidity
has led to  new analyses of factorization breaking, nonlinear mixing, event shape selection, and
forward-backward
fluctuations~\cite{Khachatryan:2015oea,Aad:2014fla,Aad:2015lwa,Adam:2015eta,Adamczyk:2015xjc,Adare:2015ctn}.
In this paper we analyze the detailed structure of two-particle transverse
momentum correlations by using  event-by-event (boost-invariant) hydrodynamics
and principal component analysis (PCA) 
\cite{Bhalerao:2014mua,Mazeliauskas:2015vea}. Specifically, we
decompose the event-by-event harmonic flow $V_n(p_T)$
into  principal components and investigate the physical
origin of each of these fluctuations. This paper extends our previous analysis~\cite{Mazeliauskas:2015vea} for
triangular flow at the LHC (Pb+Pb at $\snn=2.76\,{\rm TeV}$) to the other 
harmonics, $n=0\text{--}5$. In particular, we 
demonstrate the importance of radial flow fluctuations for subleading flows of 
higher harmonics.

Taking the second harmonic for definiteness,  the two-particle correlation
matrix of momentum dependent elliptic flows, $C_2(p_{T1},p_{T2}) \equiv \llangle V_2(p_{T1}) V_2^*(p_{T2}) \rrangle$ is traditionally parametrized by  $r_2(p_{T1}, p_{T2})$~\cite{Gardim:2012im},
\st
r_2(p_{T1}, p_{T2}) \equiv \frac{ \llangle V_2(p_{T1}) V_2^*(p_{T2}) \rrangle}{\sqrt{ \llangle |V_2(p_{T1})|^2 \rrangle
\llangle |V_2(p_{T2})|^2  \rrangle } }\,.
\stp  

If there is only one source of elliptic flow in the event [for example if
in each event
$V_2(p_T) = f(p_T) \varepsilon_2$ 
with $\varepsilon_2$ a complex eccentricity and
$f(p_T)$ a fixed real function of $p_T$]  then the correlation matrix of 
elliptic
flows $C_2(p_{T1},p_{T2})$ factorizes into a product of functions, and the
$r_2$ parameter is unity.
However, if there are multiple independent sources of elliptic flow in the 
event, then the correlation matrix does not factorize, and the 
$r_2$ parameter is less than unity~\cite{Gardim:2012im}.  
The $r_2$ parameter
 has been extensively studied 
both 
experimentally~\cite{Khachatryan:2015oea,Aamodt:2011by,ATLAS:2012at,Aad:2014lta}
 and 
theoretically~\cite{Gardim:2012im,Heinz:2013bua,Kozlov:2014fqa,Mazeliauskas:2015vea}.
 In particular, in our prior
work on triangular flow we showed that factorization breaking in event-by-event
hydrodynamics arises because the simulated triangular flow is predominantly the result of two
statistically uncorrelated contributions---the linear response to 
$\varepsilon_3$~\cite{Alver:2008aq} and the
linear response to the first radial excitation of 
$\varepsilon_3$~\cite{Mazeliauskas:2015vea}. The goal of
the current paper is to extend this understanding of factorization breaking to 
the other harmonics. This extension was surprisingly subtle due to the quadratic
mixing between the leading and subleading harmonic flows. 

Experimentally, it is observed that factorization breaking is largest for 
elliptic flow in
central collisions (see in particular Fig.~28 of Ref.~\cite{ATLAS:2012at} and 
Fig.~1 of Ref.~\cite{Khachatryan:2015oea}). Indeed, the $r_2$ parameter decreases
rather dramatically from mid-central  to central collisions. This indicates
that the relative importance of the  various initial state fluctuations which
drive elliptic flow are changing rapidly as a function of centrality.  The
current manuscript explains the rapid centrality dependence of factorization
breaking in $v_2$ as an interplay between the linear response to the
fluctuating elliptic geometry, and the nonlinear mixing of the radial
flow and average elliptic flow. 
This quadratic mixing is similar to the mixing between $v_5$ and $v_2,v_3$ 
~\cite{Borghini:2005kd,Qiu:2011iv,Gardim:2011xv,Teaney:2012ke}, and 
this picture can be confirmed experimentally by measuring specific three point 
correlations analogous  to the three plane correlations
measured in the $v_5,v_2,v_3$ case~\cite{Aad:2014fla,Aad:2015lwa}.

To understand the linear and nonlinear contributions quantitatively, we will break
up the fluctuations in hydrodynamics into their principal components, and
analyze the linear and nonlinear contributions of each principal component to
the simulated harmonic spectrum.  The sample of events and most of the PCA methods
are the same as in our previous paper~\cite{Mazeliauskas:2015vea}, and therefore 
in Sects.~\ref{principals} and \ref{simulations} we only briefly review the 
analysis definitions,  and the key features of simulations. In 
\Sect{predictors} we discuss the strategy for constructing the best linear 
predictor for leading and subleading flows.

The second part of our paper, \Sect{results}, contains individual discussions 
for each harmonic flow. First, we discuss radial flow fluctuations in 
\Sect{radial} and then demonstrate their importance in generating subleading 
elliptic flow in \Sect{elliptic}. In 
\Sect{v1v3} we briefly describe our PCA results for direct and triangular flows. 
Finally, in \Sect{v4v5} we discuss the quadrangular and pentagonal flows and 
how the nonlinear mixing of lower order principal components adds to 
these flows. 
We put forward some experimental observables in the discussion in 
\Sect{discussion}. A catalog of 
figures showing the main results of PCA for each harmonic is given in the 
\hyperref[lof]{Appendix}.

%%%%%%%%%%%%%%%%%%%%%%%%%%%%%%%%%%% pca.tex %%%%%%%%%%%%%%%%%%%%%%%%%%%%%%%%%%%%

\section{PCA of two-particle correlations in event-by-event hydrodynamics}
\label{pca}
\subsection{Principal components}
\label{principals}
PCA is a statistical technique for extracting the dominant components in 
fluctuating data. In the context of flow in heavy ion collisions it was first 
introduced in Ref.~\cite{Bhalerao:2014mua}, 
and then applied to the 
analysis of triangular flow in our previous paper~\cite{Mazeliauskas:2015vea}. 
Here we review  the essential definitions.

The event-by-event single particle distribution is customarily expanded in a 
Fourier series
\st
\frac{d N}{d \p}   = V_0(p_T) + \sum_{n=1}^{\infty} V_n(p_T) e^{-in \varphi}  + 
{\rm H.c.}\, ,\label{series}
\stp
where $d\p =  d y \,d p_T\, d\varphi$ denotes the phase space, 
$\varphi$  is the
azimuthal angle of the distribution, and H.c. denotes Hermitian 
conjugate. $V_n(p_T)$ is a complex Fourier
coefficient recording the magnitude and orientation of the $n$th harmonic 
flow, without the typical normalization by multiplicity.

PCA is done by expanding the covariance matrix 
of two-particle correlations 
(which is real, symmetric, and positive 
semi-definite) 
into real orthogonal eigenvectors $V_n^{(a)}(p_T)$,
\begin{align}
   C_n(p_{T1},p_{T2})&\equiv \big< \big(V_{n,p_{T1}}-\left< 
   V_{n,p_{T1}}\right>\big) \big( V^*_{n,p_{T2}}-\left< 
   V^*_{n,p_{T2}}\right>\big) \big> \nonumber \\
&= \sum_{a} V_n^{(a)}(p_{T1}) V^{(a)}_n(p_{T2})\label{dec2},
\end{align}
where $V_n^{(a)} = \sqrt{\lambda_a} \psi^{(a)}(p_T)$ is the square root of the eigenvalue times a normalized eigenvector. The eigenvalue records the squared variance of a given fluctuation. 

The principal components $V_n^{(1)}(p_{T}), V_n^{(2)}(p_{T}),\ldots$ of a given 
event 
ensemble 
can be used as optimal basis 
for event-by-event expansion of harmonic flow
\st
V_n(p_{T}) -\left<V_n(p_{T})\right> = \xi_n^{(1)} V^{(1)}_n(p_{T}) + 
\xi_n^{(2)} 
V^{(2)}_n(p_{T}) +\ldots  \, .
\stp
The complex coefficients $\xi_n^{(a)}$ are the projections of 
harmonic flow onto principal component basis and record  the orientation and 
event-by-event amplitude of their respective flows.
Principal components are mutually uncorrelated
\begin{equation}
\left<\xi_n^{(a)}\xi_n^{*(b)}\right> = \delta^{ab}.\label{orth}
\end{equation}

Typically the eigenvalues of $C_n(p_{T1},p_{T2})$ are strongly ordered and only the 
first 
few terms in the expansion are significant. Often the large components
have a definite physical interpretation. We define the 
scaled magnitude of the flow vector $V_n^{(a)}(p_T)$ as
\st
\label{pcamagnitudes}
\| v^{(a)}_n\|^2  \equiv \frac{\int \big(V_n^{(a)}\!(p_T)\big)^2dp_T}{\int 
\left<dN/dp_T\right>^2dp_T},
\stp
which is a measure of the size of the fluctuation without
trivial dependencies on the mean multiplicity in a given centrality class.
\begin{figure}
\centering
\includegraphics[width=\columnwidth]{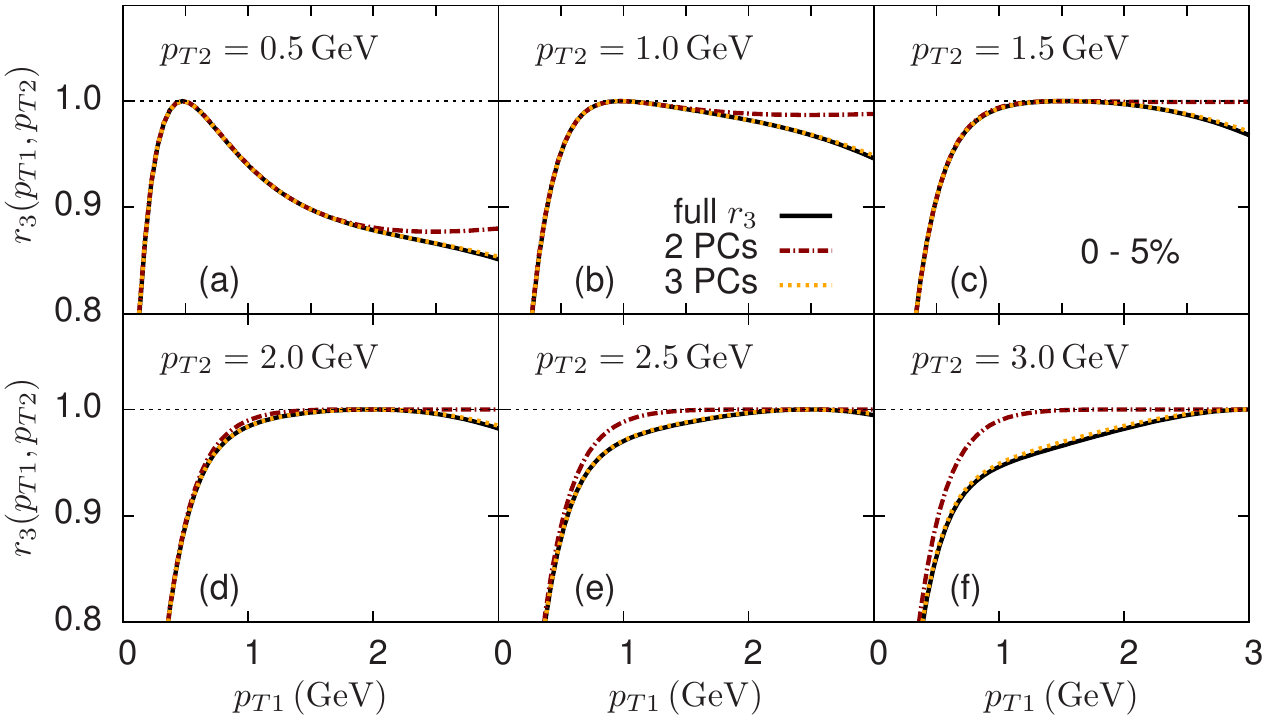}
\caption{Factorization ratio $r_3(p_{T1},p_{T2})$ [\Eq{rij}] for triangular flow and its 
approximations 
with principal components (PCs) in central collisions (0--5\%). }
\label{fig:rij}
\end{figure}

The leading flow vector $V^{(1)}_n(p_T)$ corresponds to fluctuations 
with the largest root-mean-square amplitude, while subsequent components maximize the 
variance  in the remaining orthogonal directions. This yields a very efficient 
description of the full covariance matrix $C_n(p_{T1},p_{T2})$ and 
factorization ratio
\begin{equation}
r_n(p_{T1}, p_{T2}) \equiv \frac{ C_n(p_{T1}, p_{T2})}{\sqrt{ C_n(p_{T1}, 
p_{T1}) 
C_n(p_{T2}, p_{T2})}}\leq 1\, \label{rij}.
\end{equation}
$r_n(p_{T1},p_{T2})$ is bounded by unity  within hydrodynamics~\cite{Gardim:2012im}. By truncating series expansion 
of the covariance matrix
[\Eq{dec2}] at two or three principal components we can 
approximate $C_n(p_{T1},p_{T2})$ and $r_n(p_{T1},p_{T2})$. Truncating  
at the leading term would constitute complete flow factorization, i.e., 
$r_n=1$. 
The factorization matrix for triangular flow is
shown in \Fig{fig:rij}. We see that at low momentum 
$p_T<2\,\rm{GeV}$ just two 
principal components are sufficient to describe momentum dependence of 
factorization ratio $r_3$. Analogous decompositions of two-particle correlations
into principal components exist for all harmonics and 
all centralities. 
Interpreting these large flow components physically is the goal of this paper.

\subsection{Simulations}
\label{simulations}
We used boost-invariant event-by-event viscous hydrodynamics to simulate 5000 
Pb-Pb  collisions at the CERN Large Hadron Collider (LHC) ($\snn=2.76\,{\rm 
TeV}$) in fourteen 5\% centrality classes 
selected by impact parameter.
The initial conditions are based on the Phobos Glauber Monte Carlo 
\cite{Alver:2008aq} with a two-component model for the entropy distribution in 
the transverse plane. We used a lattice equation of 
state~\cite{Laine:2006cp}
and a ``direct" pion freeze-out at $T_{\rm fo} = 140\,{\rm MeV}$. The results 
presented here were simulated using a shear viscosity to entropy ratio of 
$\eta/s=0.08$. Qualitatively similar results are obtained with $\eta/s=0.16$, 
with the most important differences discussed in \Sect{viscosity}.
The same ensemble of events was used in 
Ref.~\cite{Mazeliauskas:2015vea}, which provides further simulation details.

\subsection{Geometrical predictors}
\label{predictors}
We will construct several geometric predictors for the leading
and subleading flows following strategy outlined in Ref.~\cite{Gardim:2011xv}. Keeping the 
discussion general, let $\xi^{(a)}_{n\,\text{pred}}$ be a geometric quantity which predicts the event-by-event amplitude and phase of 
the corresponding flow $\xi^{(a)}_n$ .
For example, for the leading $n=3$ component
the triangularity $\varepsilon_{3,3}$ (defined below) is an excellent choice for $\xi_{3\,\text{pred}}^{(1)}$.

The geometric predictors are 
designed to maximize the correlation
between a particular flow signal and the geometry. Specifically,
the predictors maximize the Pearson correlation coefficient between the 
event-by-event magnitude and orientation of $a$th principal 
component, $\xi^{(a)}_n$, and the geometrical predictor $\xi^{(a)}_{n\,\text{pred}}$
\begin{align}
	\text{max}\quad	Q^{(a)}_n=\frac{\big<\xi^{(a)}_n 
	\xi^{*(a)}_{n\,\text{pred}}\big>}{\sqrt{\big<\xi^{(a)}_n 
	{\xi^*_n}^{(a)}\big>\big<\xi^{(a)}_{n\,\text{pred}}\xi^{*(a)}_{n\,\text{pred}}\big>}}.\label{pears}
\end{align}

We constructed several predictors 
for the flow $\xi^{(a)}_n$ by 
assuming a linear relation between the flow and the geometry.
The simplest predictor consists of linear combinations of the
first two eccentricities of the initial geometry. These are defined as 
\begin{subequations}
	\label{eps3all}
	\begin{align}
		\varepsilon_{n,n} &\equiv   -\frac{[r^n e^{in \phi }]}{ R_{\rm rms}^n } 
		\, , \label{eps33} \\
		\varepsilon_{n,n+2} &\equiv   -\frac{ [r^{n+2} e^{in \phi }]}{ R_{\rm 
		rms}^{n+2}}  
		\, ,\label{eps35} 
	\end{align} %\label{eps3} 
\end{subequations} 
where the square brackets $[\,]$ denote an integral over the initial entropy 
density for a specific event, normalized by the average total entropy 
$\bar{S}_\text{tot}$.  $R_{\rm rms}\equiv\sqrt{\llangle[r^2]\rrangle}$ is 
the event averaged root-mean-square radius. Note that our definitions of
$\varepsilon_{n,n}$ and $\varepsilon_{n,n+2}$ are chosen to make 
the event-by-event quantities $\varepsilon_{n,n}$ and $\varepsilon_{n,n+2}$ 
linear in the fluctuations. In this notation, the geometric predictor based on 
these eccentricities is
\begin{equation}
\xi_{n\,\text{pred}}^{(a)} = \varepsilon_{n,n} + c_1 \, \varepsilon_{n,n+2},
\end{equation}
where $c_1$ is adjusted to maximize the correlation coefficient in 
\Eq{pears}, 
and 
the overall normalization is irrelevant.
While the first two eccentricities provide an excellent predictor for
the leading flow, they do not predict  the subleading flow very 
well. This is  in part because the radial weight $r^{n+2}$ is too 
strong at large $r$.

More generally, one can define eccentricity as a functional of radial weight 
function $\rho(r)$:
\begin{equation}
\varepsilon_n\{\rho(r)\} \equiv - \frac{[\rho(r) 
e^{in\phi}]}{R_\text{rms}^n}.\label{epsgen}
\end{equation}
It is the goal of this paper to find the optimal radial weight
function $\rho(r)$ for predicting both leading and subleading flows. 
For the subleading modes $\rho(r)$ will have a node, 
and thus  $\varepsilon_n\{\rho(r)\}$ will measure the magnitude and
orientation of the first radial excitation of the geometry~\cite{Mazeliauskas:2015vea}.

To find the optimal radial weight we expand $\rho(r)$ in radial Fourier modes
\begin{equation}
\rho(r)= \sum_{b=1}^{n_k} w_b \frac{2^n n!}{k_b^n} J_n(k_b r),\label{exp}
\end{equation}
where $J_n(x)$ is a Bessel function of order $n$, $w_b$ are expansion
coefficients, and $k_b$ are definite wavenumbers  specified below.
The prefactor is chosen 
so that for a single $k$ mode ($w_1=1,w_{b>1}=0$) at small $k$ ($k R_{\rm rms} 
\ll 1$) the 
generalized
eccentricity approaches $\varepsilon_{n,n}$ 
\st
\lim_{k \rightarrow 0} \varepsilon_n\{\rho(r)\} = \varepsilon_{n,n}.
\stp
At small $k$, we expand the $J_{n}(k r)$ and find
\st
    \varepsilon_n\{\rho(r)\} \simeq \varepsilon_{n,n}  + c_1 \, 
    \varepsilon_{n,n+2}, 
    \stp
    where $c_1 = -(k R_{\rm rms}/2)^2/(1+n)$.
Thus the functional form of $\rho(r)$  adopted here
 yields a tunable linear combination of the eccentricities in \Eq{eps3all},
 but the wave number 
parameter regulates the behavior at large $r$.
Further motivation and discussion of this basis set for 
$\rho(r)$ is given in our previous work~\cite{Mazeliauskas:2015vea}.

We have found that an approximately optimal radial weight
can be found by using only two well chosen 
$k_b$ values for the Fourier expansion in \Eq{exp}. 
Including additional $k$ modes in the functional
form of $\rho(r)$ does not significantly improve the predictive power
of the generalized geometric eccentricity.
For the two $k$ modes we required (somewhat
arbitrarily) that the ratio of $k$ values would 
be fixed to the ratio of the first two Bessel zeros:
\begin{equation}
\frac{k_1}{k_2}=\frac{j_{n,1}}{j_{n,2}}\label{j0}  \, .
\end{equation}
With this choice our basis functions were orthogonal in the interval 
$[0,R_o]$,  where $k_1 = j_{n,1}/R_o$.   We then adjusted $R_o$ to maximize the 
correlation coefficient 
between $\varepsilon_{n}\{\rho(r)\}$ and the flow $\xi^{(a)}_n$. To account for 
changing system size with centrality, we used a fixed
$R_o/R_\text{rms}$ ratio. In most cases we used $R_o/R_\text{rms}\approx 3.0$, 
but for all directed flow components ($\xi_1^{(1)}$ and $\xi_1^{(2)}$) and the second elliptic flow component ($\xi_2^{(2)}$), we found that 
$R_o/R_\text{rms}\approx 2.0$ optimized the correlation between the flow and 
the geometry.

Ultimately, the assumption that the amplitude and phase of the flow is 
determined at least approximately by initial eccentricity,
$\varepsilon_{n}\{\rho(r)\}$,  
is based on linear response.  
If nonlinear physics becomes important
(as in the case of $v_4$ and $v_5$) then the predictors should be 
modified to incorporate this physics (see below and Ref.~\cite{Gardim:2011xv}). Thus, below
we will refer to the $\varepsilon_{n}\{\rho(r)\}$ (with an optimized
radial weight) as the \emph{best linear predictor} and incorporate
quadratic nonlinear corrections to the predictor as needed.

%%%%%%%%%%%%%%%%%%%%%%%%%%%%%%%%% results.tex %%%%%%%%%%%%%%%%%%%%%%%%%%%%%%%%%%

\section{Results}
\label{results}

\subsection{Radial flow}
\label{radial}

Radial flow (or $V_0(p_{T})$) is the first term in the Fourier series  and 
is 
by far the largest harmonic. Traditionally, the experimental and theoretical study of the fluctuations of 
$V_0(p_{T})$ (i.e., multiplicity and $p_{T}$ fluctuations) has been distinct 
from elliptic and triangular flow. There is no reason for this distinction.

\begin{figure*}
\centering
\subfigure[]{\includegraphics{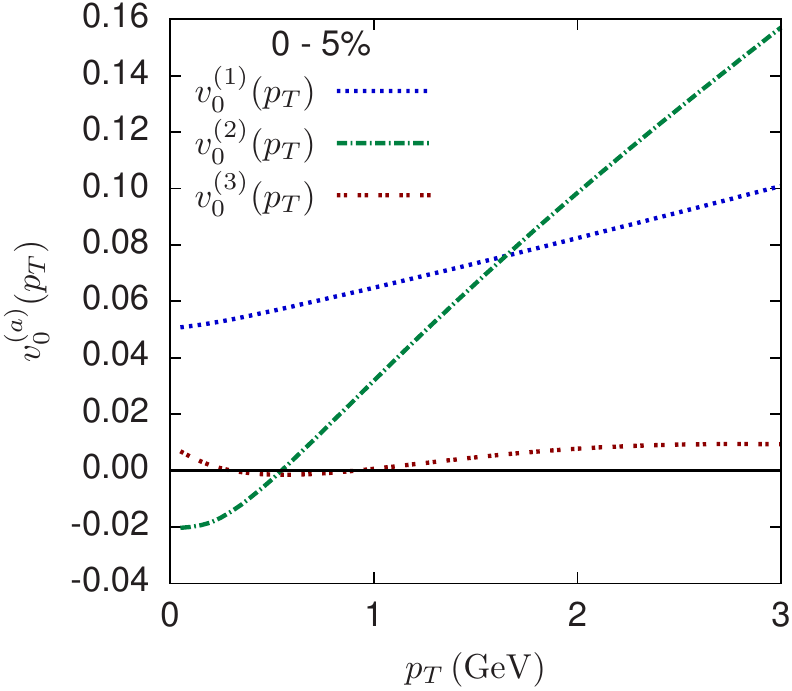}}
\subfigure[]{\includegraphics{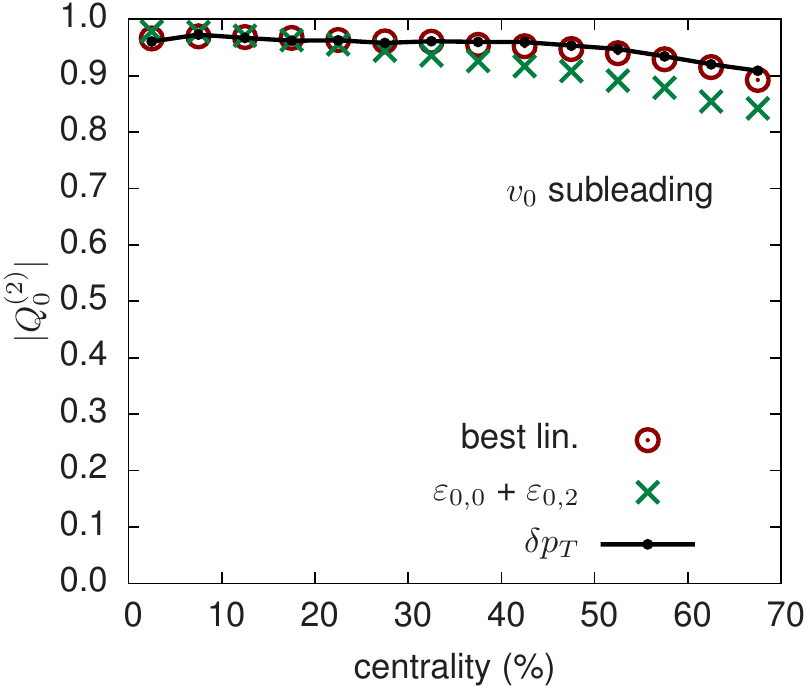}}
\caption{(a) The $p_T$ dependence of the principal components of radial flow normalized by the average multiplicity, 
$v_0^{(a)}(p_T)\equiv 
V_0^{(a)}(p_T)/\left<dN/dp_T\right>$. (b) The Pearson correlation coefficient [\Eq{pears}]  between
the subleading radial flow and various predictors versus centrality. The best linear predictor is described in \Sect{predictors}.\label{v0-fig}}
\end{figure*}

Examining the scaled $V_0(p_T)$ eigenvalues shown in \Fig{v0-fig}(a), we see 
that there are
two large  principal components.
The first principal component is sourced by multiplicity fluctuations,
i.e., the magnitude of $V_0(p_{T})$ fluctuates (but not its shape)
due to the impact parameter variance in a given centrality bin.  
Corroborating this inference,
\Fig{v0-fig}(a) shows the momentum dependence of the leading principal 
component, which is approximately flat.\footnote{There is a small upward 
tending slope in our
simulations of this component, because multiplicity and mean $p_T$ fluctuations 
only approximately factorize into leading and subleading principal components. 
Using different definitions of centrality bins could perhaps make this separation cleaner.
} Clearly this principal component is not
particularly interesting, and the PCA procedure gives a practical method for isolating these trivial geometric fluctuations in the data set.
The second principal component is of much greater interest, and shows a linear  
rise with $p_T$ that  is  indicative of the fluctuations in the radial
    flow velocity of the fluid~\cite{Bhalerao:2014mua}.

In  early insightful papers~\cite{Broniowski:2009fm,Bozek:2012fw}, the 
fluctuations in the flow velocity (or mean $p_T$) 
were associated with the fluctuations in the initial fireball radius. 
These radial fluctuations are well described by both the
eccentricities $(\varepsilon_{0,0}$, $\varepsilon_{0,2})$, \Eq{eps3all}, and the 
optimized 
eccentricity $\varepsilon_{0}\{\rho(r)\}$, \Eq{epsgen}.   Therefore,
as seen in \Fig{v0-fig}(b),
the subleading flow signal is strongly correlated with these linear
geometric predictors.

Also shown in  \Fig{v0-fig}(b) is the correlation between subleading 
radial flow $\xi_0^{(2)}$ and mean transverse momentum fluctuations around the average
\begin{equation}
\delta p_T\label{dp}\equiv [p_T] - \left<[p_T]\right>.
\end{equation}
Indeed, the subleading radial flow correlates very well with mean momentum 
fluctuations in all centrality bins.

In the next sections we will study the nonlinear mixing between the radial flow
$\xi_0^{(2)}$ and all other harmonics.

\subsection{Elliptic flow}

\subsubsection{Nonlinear mixing and elliptic flow}
\label{elliptic}

\begin{figure}
\centering
\includegraphics{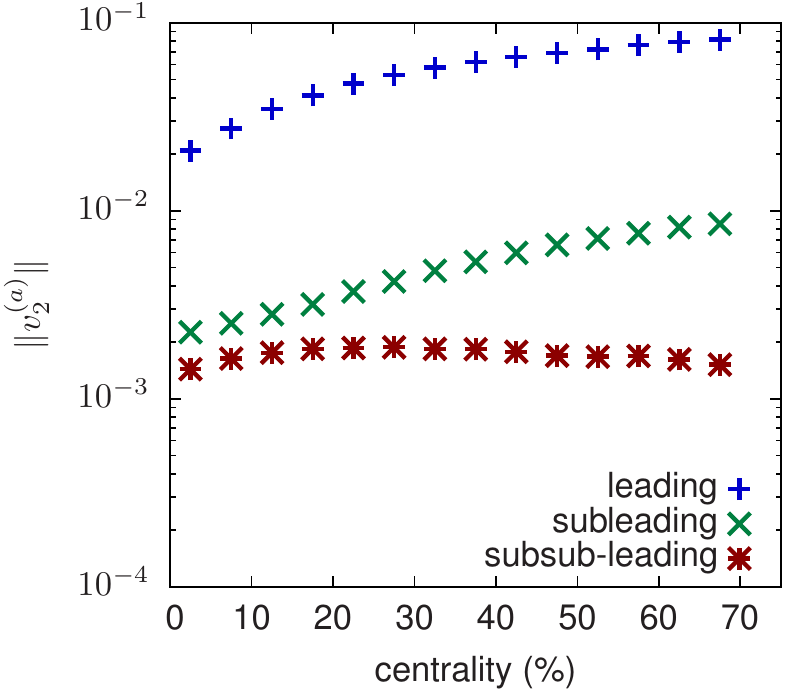}
\caption{The magnitudes of the principal
components of elliptic flow, $\|v^{(a)}_2\|$, versus centrality [see 
\Eq{pcamagnitudes}].
\label{pv2-eval}}
\end{figure}
\begin{figure*}
\centering
\subfigure[]{\includegraphics{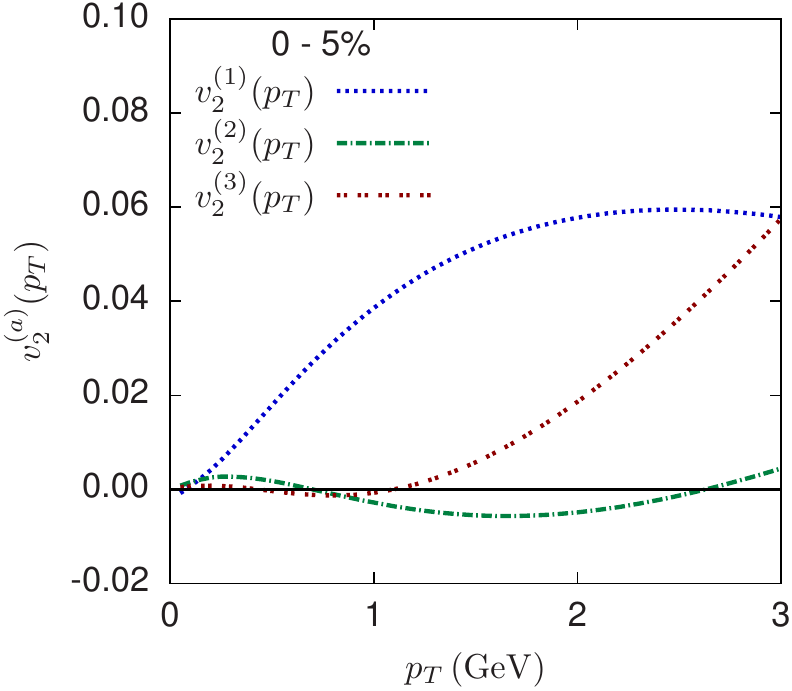}}
\subfigure[]{\includegraphics{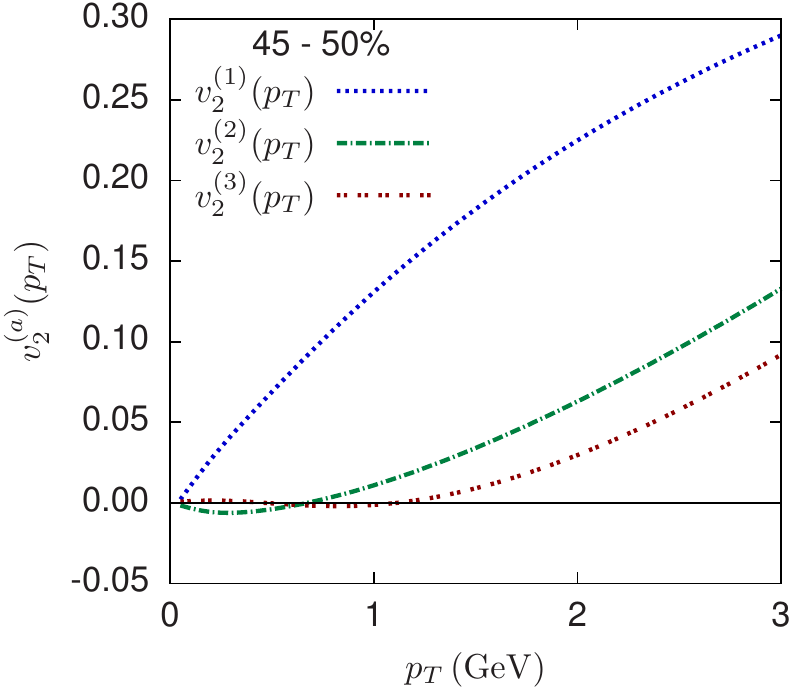}}
\caption{The $p_T$ dependence of the principal components of elliptic flow normalized by the average multiplicity, 
$v_2^{(a)}(p_T)\equiv 
V_2^{(a)}(p_T)/\left<dN/dp_T\right>$, for central (0--5\%) and peripheral
collisions (45--50\%). 
\label{v2-vn_norm-p}
}
\end{figure*}

We next study the fluctuations of $V_2(p_T)$ as function of centrality.
As seen in \Fig{pv2-eval}, the principal component spectrum of elliptic flow 
in 
central collisions 
consists of two nearly degenerate subleading 
components in addition to the dominant leading component. This degeneracy is lifted in more peripheral 
bins. Comparing the $p_T$ dependence of the principal flows
shown in \Figs{v2-vn_norm-p}(a) and \ref{v2-vn_norm-p}(b), we see that 
going 
from 
central (0\nobreakdash--5\%) to peripheral (45\nobreakdash--50\%) collisions, 
the magnitude of the second principal component  increases 
in size and its momentum dependence changes
dramatically. By contrast, the growth of the third principal component is
much more mild. This strongly suggests that the average elliptic geometry
is more important for the subleading  than the subsub-leading mode.

\begin{figure*}
\centering
\subfigure[]{\includegraphics{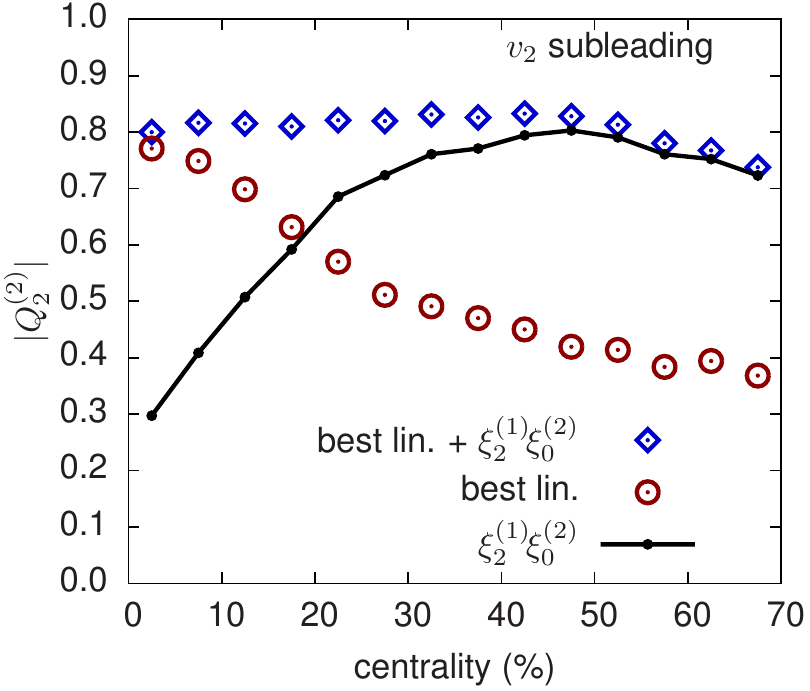}}
\subfigure[]{\includegraphics{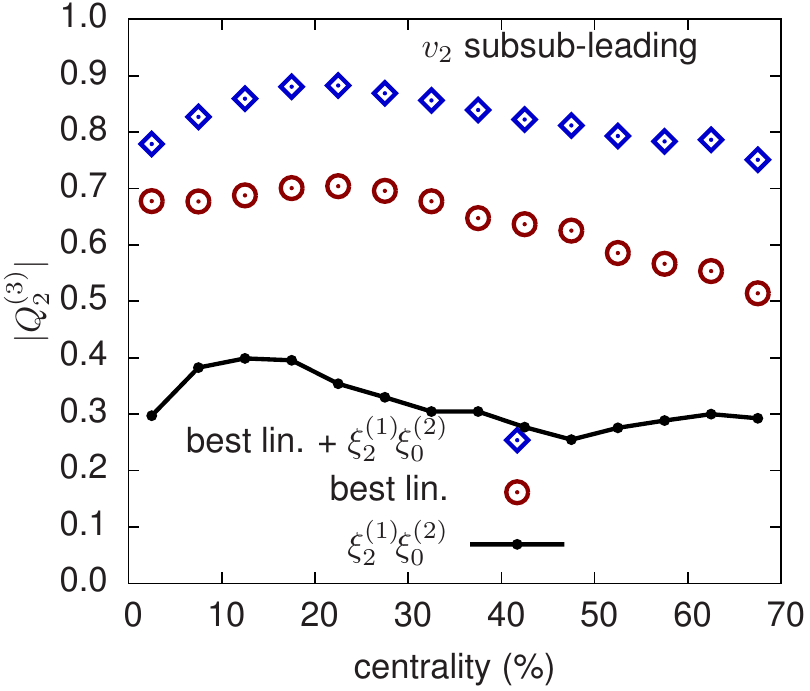}}
\caption{
    Pearson correlation coefficient 
    between the subleading elliptic flows and the best linear predictor [\Eq{epsgen}] 
    with and without the nonlinear
    mixing between the radial and leading elliptic flows, 
    $\xi_2^{(1)}\xi_0^{(2)}$. (a) and (b) show the correlation 
    coefficient for $v_2$ subleading and $v_2$ subsub-leading flows 
    respectively.
\label{pv2-corr}
}
\end{figure*}

To find a geometrical predictor for the sub- and sub-sub-leading modes
we first tried the best linear predictor $\varepsilon_2\{\rho(r)\}$. 
In \Fig{pv2-corr}(a) (the red circles), we see that the correlation coefficient between this optimal linear predictor and the subleading flow signal drops precipitously as a function
of centrality. As we will explain now, this is because nonlinear mixing
becomes important for the subleading mode.

The ellipticity of the almond shaped geometry in peripheral collisions is 
traditionally parametrized by eccentricity $\varepsilon_{2,2}$
and it serves as an excellent predictor for the leading elliptic flow.
However, $\varepsilon_{2,2}$ does not completely fix the 
initial geometry, and the radial size of the fireball can fluctuate at fixed 
eccentricity. As 
explained in \Sect{radial}, the radial size fluctuations 
modulate the momentum spectrum of the produced particles, and for a background 
geometry with large {\em constant} eccentricity 
this generates
fluctuations in the $p_T$ dependence of the elliptic flow, i.e., subleading 
elliptic flow. This subleading flow lies in the reaction plane following the
average elliptic flow, but its sign (which is determined by $\delta p_T$) is 
uncorrelated with $\varepsilon_{2,2}$.

The orientation of the 
reaction plane in peripheral bins is strongly correlated with the 
integrated 
$v_2$ or the 
leading elliptic principal component $\xi_2^{(1)}$, while the mean 
$p_T$ fluctuations are tracked by the subleading radial flow component $\xi_0^{(2)}$. Therefore we 
correlated the sub- and sub-sub-leading  elliptic flows
with the 
product of the  leading elliptic and radial flows, i.e. 
we computed the correlation coefficient in \Eq{pears} with
$\xi_{2,\rm pred}^{(2)}=\xi_2^{(1)}\xi_0^{(2)}$. 
Examining \Fig{pv2-corr}(a) (the black line), we see 
see that 
the correlation between the subleading elliptic flow and
the nonlinear mixing rises with
centrality,  as the correlation with best linear predictor drops. Examining \Fig{pv2-corr}(b) on the other hand, we see that the 
subsub-leading elliptic flow has 
stronger correlation with the initial geometry than the nonlinear mixing.  Combining 
best linear geometric predictor and quadratic mixing terms  in the 
predictor, i.e.
\st
\xi_{2\,\rm pred}^{(2)}  = \varepsilon_{2}\{\rho(r)\} + c_1 
\xi_2^{(1)}\label{pred2} 
\xi_0^{(2)}, 
\stp
we achieve consistently 
high correlations for all centralities [the blue diamonds in \Fig{pv2-corr}(a) 
and (b)].

\subsubsection{Dependence on viscosity}
\label{viscosity}

\begin{figure}
\centering
\includegraphics{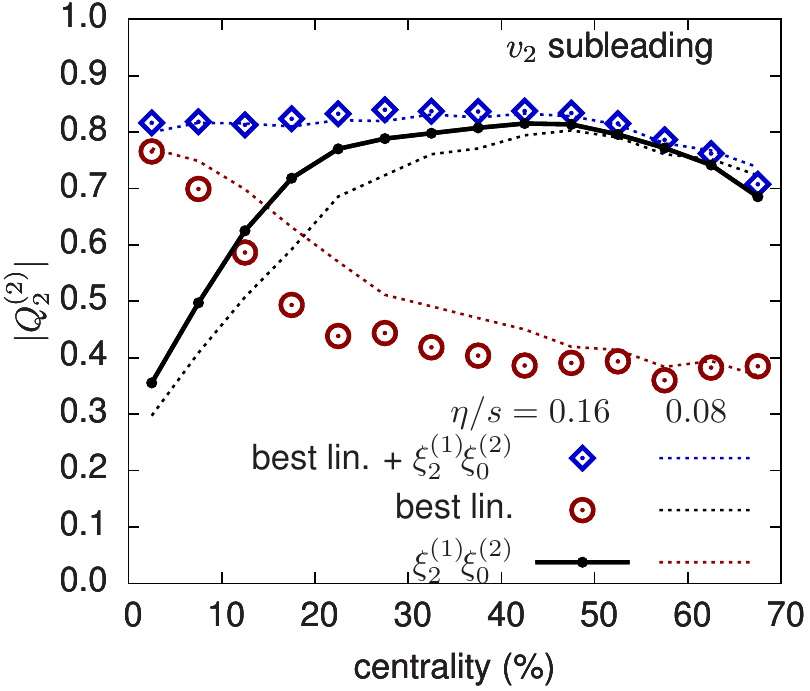}
\caption{
     Pearson correlation coefficients  for the subleading elliptic flow at 
     viscosity over entropy ratio $\eta/s=0.16$. Dashed lines repeat 
     $\eta/s=0.08$ results from \Fig{pv2-corr}(a) for the ease of comparison.
\label{pv2s16-corr}
}
\end{figure}

Before leaving this section we will briefly comment  on the viscosity
dependence of these results. Figure \ref{pv2s16-corr} shows a typical 
result  for
a slightly larger shear viscosity, $\eta/s=0.16$.
As discussed above, the subleading elliptic 
flow [i.e., the event-by-event fluctuations in $V_2(p_T)$] is a result of the
linear response to the first radial excitation of the elliptic eccentricity,
and a nonlinear mixing of radial flow fluctuations and the leading elliptic
flow. In \Fig{pv2s16-corr} we see that a slightly larger shear viscosity 
tends to preferentially damp the linear response leaving a stronger nonlinear
signal. This is because
the initial geometry driving the linear response has a significantly larger 
gradients due to the combined azimuthal and radial variations. Thus in 
\Fig{pv2s16-corr} the linear response  dominates the 
subleading flow only in 
very central collisions.
These trends with centrality are qualitatively familiar from previous analyses of
the effect of shear viscosity on the nonlinear mixing of
harmonics~\cite{Teaney:2012ke,Qiu:2012uy}.

\subsection{Triangular and directed flows}
\label{v1v3}
\begin{figure*}
\centering
\subfigure[]{\includegraphics{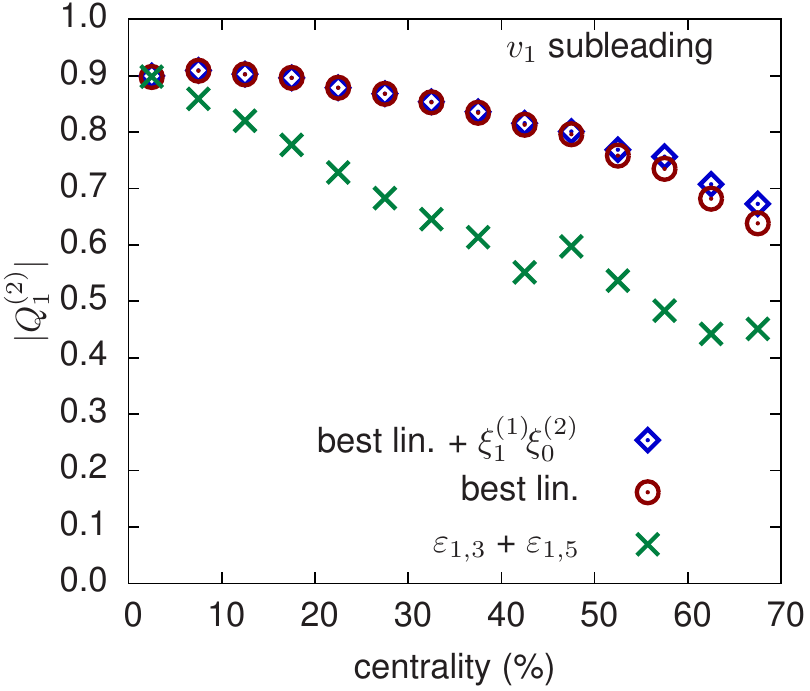}}
\subfigure[]{\includegraphics{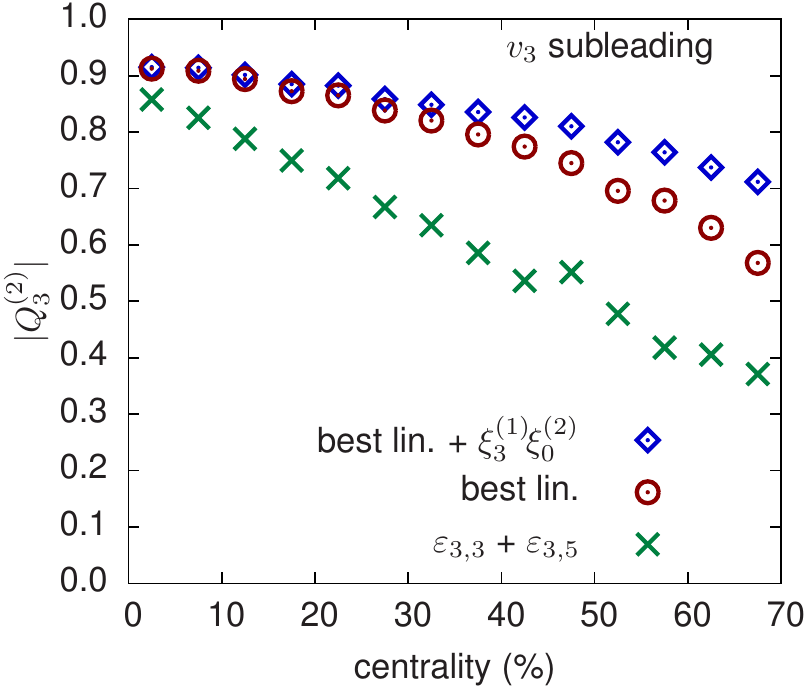}}
\caption{
    Pearson correlation coefficient 
    between the subleading (a) directed and (b) triangular flows and the best 
    linear predictor
    with and without radial flow mixing. 
\label{pv1v3-corr}}
\end{figure*}

Triangular flow  was 
extensively studied in our previous work~\cite{Mazeliauskas:2015vea}. 
For the sake of completeness we relegate several comparative plots 
to the \hyperref[lof]{Appendix}. Previously,  we constructed an
optimal linear predictor $\varepsilon_{3}\{\rho(r)\}$ for the 
subleading triangular mode 
which 
characterizes the radially excited triangular geometry. 
As shown in \Fig{pv1v3-corr}(b), adding the nonlinear mixing term $\xi_{0}^{(2)}\xi_3^{(1)}$ to the best linear predictor 
marginally improves the already good correlation with the subleading flow  in peripheral collisions.

Directed flow exhibits many similarities to triangular flow.  Specifically
the subleading directed flow is reasonably well correlated with the optimal
linear predictor, characterizing the radially excited dipolar geometry.  Nonlinear mixing
between the  leading directed flow and the radial flow is unimportant [see 
\Fig{pv1v3-corr}(a)].

\subsection{The $n=4$ and $n=5$ harmonic flows}
\label{v4v5}
\begin{figure*}
\centering
\subfigure[]{\includegraphics{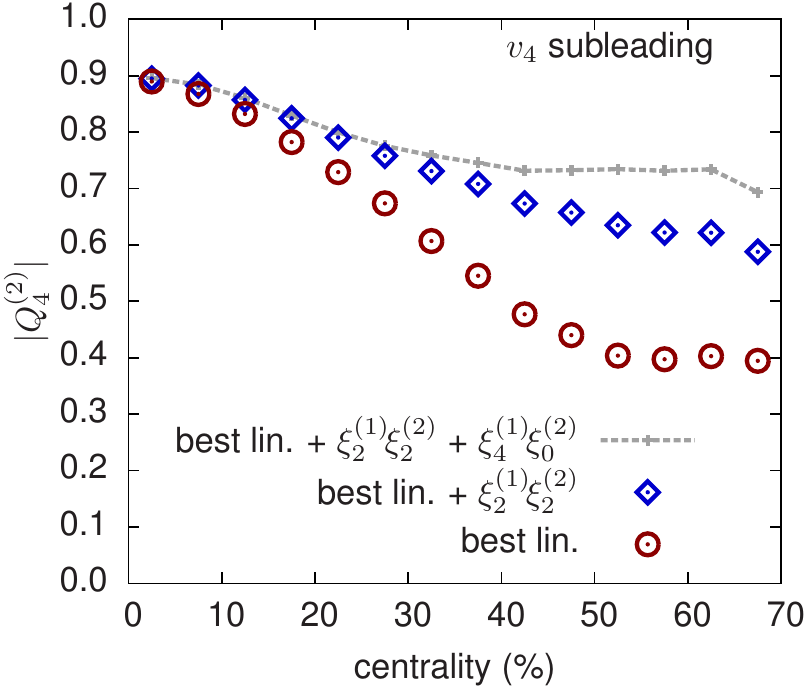}}
\subfigure[]{\includegraphics{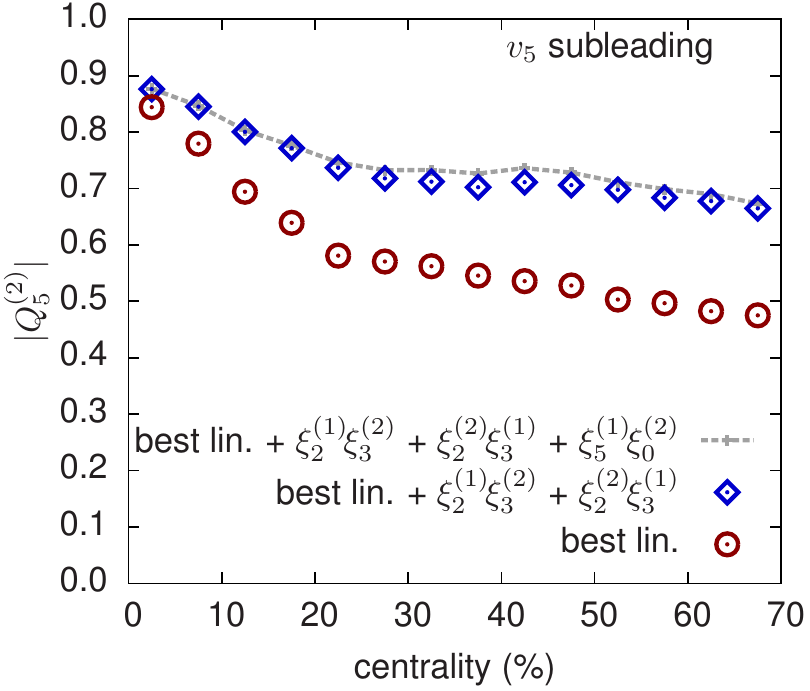}}
\caption{
    Pearson correlation coefficient 
    between  the subleading $v_4$ and $v_5$ flows and the best linear 
    predictor  with and without several nonlinear terms [see 
    \Eqs{subv4v5_predictor1} and \eq{subv4v5_predictor2}].
\label{pv4v5-corr}}
\end{figure*}

It is well known
that the leading components of the $n=4$ and $n=5$ harmonics
are determined by the nonlinear mixing of lower order harmonics in peripheral collisions~\cite{Borghini:2005kd,Qiu:2011iv,Gardim:2011xv,Teaney:2012ke,Aad:2015lwa}.

For comparison with other works~\cite{Gardim:2011xv,Gardim:2014tya}, in the 
Appendix in \Figs{v4-corr1} and \ref{v5-corr1} we 
construct a predictor based on
a linear combination of the eccentricities
\begin{align}
\varepsilon_{4,4} + c_1 \varepsilon_{2,2}\varepsilon_{2,2}\quad\text{for }n=4,\\
\varepsilon_{5,5}+ c_1 \varepsilon_{2,2}\varepsilon_{3,3}\quad\text{for }n=5,
\end{align}
where here and below the coefficient $c_1$ is adjusted to 
maximize  the correlation with the flow.
This predictor is compared
to a  linear combination of the 
optimal eccentricity $\varepsilon_n\{\rho(r)\}$ and 
the corresponding nonlinear mixings of the leading principal components
\begin{subequations}
    \label{leadingv4v5}
\begin{align}
    \varepsilon_{4}\left\{\rho(r)\right\} + c_1 \xi_2^{(1)}\xi_2^{(1)} \quad\text{for }n=4,\\
    \varepsilon_{5}\left\{\rho(r)\right\} + c_1 \xi_2^{(1)}\xi_3^{(1)}\quad\text{for }n=5.
\end{align}
\end{subequations}
Both sets of predictors perform reasonably well, though the second set has
a somewhat stronger correlation with the flow.

Returning to the subleading components, 
we first correlated with the best linear
predictors, $\varepsilon_{4}\{\rho(r)\}$ and $\varepsilon_5\{\rho(r)\}$, 
with the corresponding subleading flow signals. As seen in \Fig{pv4v5-corr} (the red circles), the correlation decreases rapidly with centrality, especially for $v_4$.  
Motivated by \Eq{leadingv4v5} which predicts the event-by-event leading $v_4$ and $v_5$ in terms $v_2$ and $v_3$, we construct a predictor for the 
subleading $v_4$ and $v_5$ in terms of the fluctuations of $v_2$ and $v_3$ (see 
Secs.~\ref{elliptic} and \ref{v1v3}, respectively). The full predictor reads
\begin{subequations}
    \label{subv4v5_predictor1}
\begin{align}
    &\varepsilon_{4}\left\{\rho(r)\right\} + c_1 \xi_2^{(1)}\xi_2^{(2)}, & \text{for }n=4,\\
    &\varepsilon_{5}\left\{\rho(r)\right\} + c_1 \xi_2^{(1)}\xi_3^{(2)} + c_2 \xi_3^{(1)}\xi_2^{(2)}, &\text{for }n=5.\label{v5nl}
\end{align}
\end{subequations}
Including the mixings between the  subleading $v_2$ and $v_3$ and the corresponding leading components greatly improves the correlation in mid-central bins (the blue diamonds).
Finally, in an effort to improve the  $v_4$ predictor in the most
peripheral bins we have added additional nonlinear mixings between
the radial flow and the leading principal components
\begin{subequations}
\label{subv4v5_predictor2}
\begin{align}
    &\varepsilon_{4}\left\{\rho(r)\right\} + c_1 \xi_2^{(1)}\xi_2^{(2)} + c_2 \xi_{4}^{(1)} \xi_0^{(2)}, & \text{for }n=4,\\
    &\varepsilon_{5}\left\{\rho(r)\right\} + c_1 \xi_2^{(1)}\xi_3^{(2)} + c_2 \xi_3^{(1)}\xi_2^{(2)} + c_3 \xi_5^{(1)} \xi_0^{(2)}, &\text{for }n=5.
\end{align}
\end{subequations}
As seen in \Fig{pv4v5-corr}(a) (the grey line) the coupling
to the radial flow improves the  correlation between the subleading $v_4$  and the
predictor in peripheral collisions. 
On the other hand, for $v_5$, \Fig{pv4v5-corr}(b), 
all of the information about the coupling to the radial flow is already included in \Eq{v5nl} and adding $v_0$ does not improve the correlation.

%%%%%%%%%%%%%%%%%%%%%%%%%%%%%%%% discussion.tex %%%%%%%%%%%%%%%%%%%%%%%%%%%%%%%%

\section{Discussion}
\label{discussion}
 In this paper  we classified the event-by-event fluctuations of the momentum 
 dependent Fourier harmonics $V_n(p_T)$ for $n=0\text{--}5$ by performing a 
 principal component analysis of the two-particle correlation matrix in 
 hydrodynamic simulations of heavy ion collisions.
The \emph{leading} principal component for each harmonic is very strongly 
correlated 
with the  integrated flow,
and therefore this component
 is essentially the familiar $v_n(p_T)$  measured in the event plane. 
The \emph{subleading} components describe 
additional $p_T$ dependent fluctuations of the magnitude and phase of $v_n(p_T)$.
This paper focuses on 
the physical origins of the subleading flows, which are the largest source of
factorization breaking in hydrodynamics.

Our systematic study started by placing radial flow (the $n=0$ harmonic) in 
the same framework as the other harmonic flows. We identified the subleading 
$n=0$ principal component  with mean $p_T$  fluctuations and confirmed (as is well known~\cite{Broniowski:2009fm,Bozek:2012fw})
that these fluctuations are predicted by the variance of the radial size of
the fireball.

The subleading directed and triangular flows 
were shown to be a linear response to the radial 
excitations of the corresponding eccentricity of the initial geometry. 
In these cases a generalized eccentricity $\varepsilon_{n}\{\rho(r)\}$ 
with an optimized  radial weight (describing the radial excitation) 
provides a good predictor for the subleading flows (\Fig{pv1v3-corr}). 
This extends our previous analysis of $v_3$ to $v_1$~\cite{Mazeliauskas:2015vea}. 

Next, we investigated the nature of the subleading elliptic flows. 
The 
principal component analysis reveals  that in central collisions there are 
two comparable sources of subleading elliptic flow,  
but they have strikingly 
different centrality dependence (see \Figs{pv2-eval} and \ref{v2-vn_norm-p}).
In mid-peripheral collisions 
 the first subleading component 
mainly reflects
a nonlinear mixing between 
elliptic and radial flows,
and  this component is only weakly correlated with 
the radially excitations of the elliptic geometry.
The second subleading component in this centrality range is substantially smaller and more closely reflects the radial excitations.
In more central collisions, however, the nonlinear mixing with the average elliptic 
flow becomes small, and the sub and subsub-leading principal components
become comparable in size.
Thus, the rapid centrality dependence of factorization
breaking in $v_2$ is the result of an interplay between the linear response to the
fluctuating elliptic geometry, and the nonlinear mixing of the radial
and average elliptic flows. 

This nonlinear mixing can be confirmed experimentally by measuring 
the correlations between the principal components 
$\llangle \xi_{2}^{(2)} (\xi_2^{(1)} \xi_0^{(2)})^* \rrangle$ 
which is predicted in \Fig{pv2-corr}.  
The prediction is that
three point correlation between
the subleading elliptic event plane, the mean $p_T$ fluctuations, and 
the leading elliptic event plane defined by the $Q_2$ vector, i.e.,
\st
       \frac{\llangle \xi_2^{(2)} \delta p_T Q_2^* \rrangle }{ \sqrt{\llangle (\delta p_T)^2 \rrangle \llangle |Q_2|^2 \rrangle } },
\stp
changes rapidly from central to midperipheral collisions.
This correlation is analogous to 
the three plane correlations such as 
$\llangle V_5 (V_2 V_3)^* \rrangle$
measured previously~\cite{Aad:2015lwa}. 

Finally, we studied factorization breaking in $v_4$ and $v_5$. 
With the comprehensive understanding of the fluctuations of $v_2$ and $v_3$
described above, the corresponding fluctuations in $v_4$ and $v_5$ were
naturally explained as the nonlinear mixing of subleading  $v_2$ and $v_3$ 
with their leading counterparts, together with linear response to the 
quadrangular and pentagonal geometries (see \Fig{pv4v5-corr}). 

The study of the fluctuations in the harmonics spectrum presented here
shows the power of the  principal component method in elucidating the physics which drive the event-by-event flow. We hope that this motivates
a comprehensive experimental program measuring the principal components and their correlations for $n=0-5$. Such an analysis would clarify the initial state in typical and ultra-central events with unprecedented precision,  and would strongly constrain the dynamical response of the quark gluon plasma.

\begin{acknowledgments}
We thank Jean-Yves Ollitrault, Wei Li, and Jiangyong Jia for continued 
interest.  We
especially thank Soumya Mohapatra for simulating hydro events.  This work was 
supported by the U.S. Department of Energy under Contract No. DE-FG-88ER40388
\end{acknowledgments}

\bibliography{master}

%%%%%%%%%%%%%%%%%%%%%%%%%%%%%%%%% appendix.tex %%%%%%%%%%%%%%%%%%%%%%%%%%%%%%%%%

\appendix
\section*{Appendix: List of figures \label{lof}}
Here we present a comprehensive catalog of plots for each harmonic 
$n=0\text{-}5$.
Centrality dependence of flow magnitudes for $n=0$, appearing as \Fig{pv2-eval} 
in the text above,  is repeated here as \Fig{v0-eval}, and analogous plots for 
other harmonics are given in
Figs.~10-14(a).
The $p_T$ dependence of normalized principal components for radial and elliptic 
flows in central (0-5\%) collisions shown in \Figs{v0-fig}(a) and 
\ref{v2-vn_norm-p}(a) are reproduced as 
\Figs{v0-vn_norm} and \ref{v2-vn_norm} and complemented with 
\Figs{v1-vn_norm} and 12-14(c). Additionally, 
Figs.~9-14(b)
depict the same principal components, but without 
normalization by average multiplicity $\left<dN/dp_T\right>$. Finally, in the 
paper we 
showed the Pearson correlation coefficients for the subleading flows for each 
harmonic $n=0\text{-}5$ in
\Figs{v0-fig}(b),\ref{pv2-corr}(a),\ref{pv1v3-corr}(a), \ref{pv1v3-corr}(b), 
\ref{pv4v5-corr}(a) and \ref{pv4v5-corr}(b), while in this appendix we show 
results for both leading and subleading flows in 
the series of figures  Figs.~9-14(d) and Figs.~9-14(e).

\newcommand{\FLOWN}{0}
\begin{figure*}
\centering
\subfigure[
{\,Centrality dependence of the (scaled) magnitudes of flows  
$\|v^{(a)}_\FLOWN\|$.}\label{v\FLOWN-eval}]{
\includegraphics{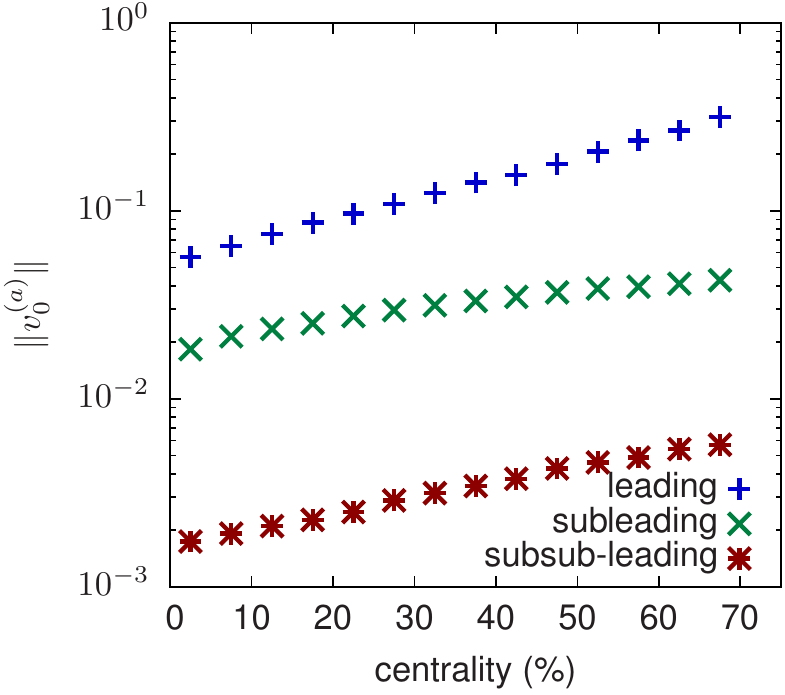}}\\
\subfigure[{\,Momentum dependence of principal flow vectors $V_\FLOWN^{(a)}(p_T)$ 
in 
central collisions.}\label{v\FLOWN-Vn_nonorm}]{
\includegraphics{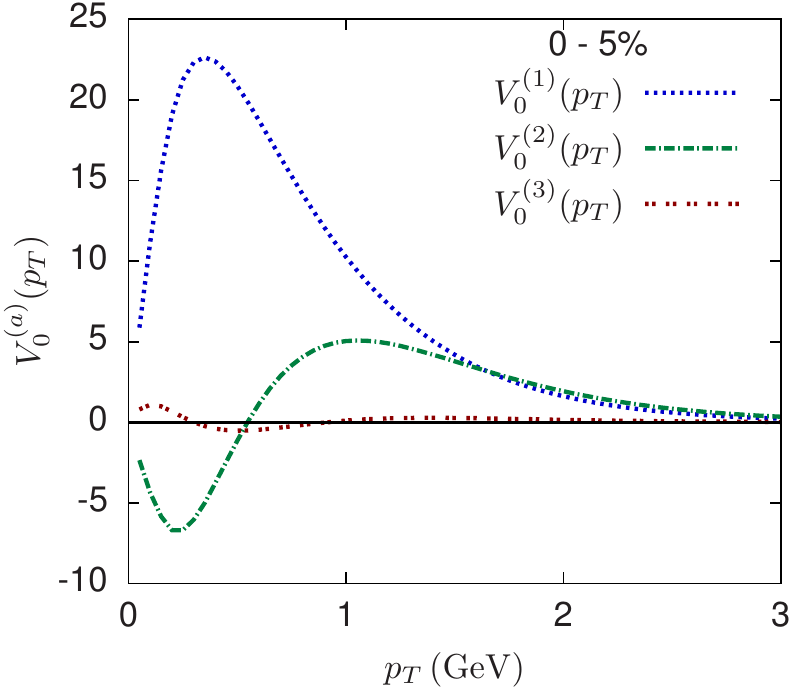}}
\subfigure[{\,Principal flow vectors divided by the average multiplicity, 
$v_\FLOWN^{(a)}(p_T)\equiv 
V_\FLOWN^{(a)}(p_T)/\left<dN/dp_T\right>$.}\label{v\FLOWN-vn_norm}]{
\includegraphics{std_A0_a0_B0_b0_CENT00_vn_norm}}
\subfigure[{\,Pearson correlation coefficient between the leading flow (zero 
suppressed for clarity) and several predictors.}\label{v\FLOWN-corr1}]{
\includegraphics{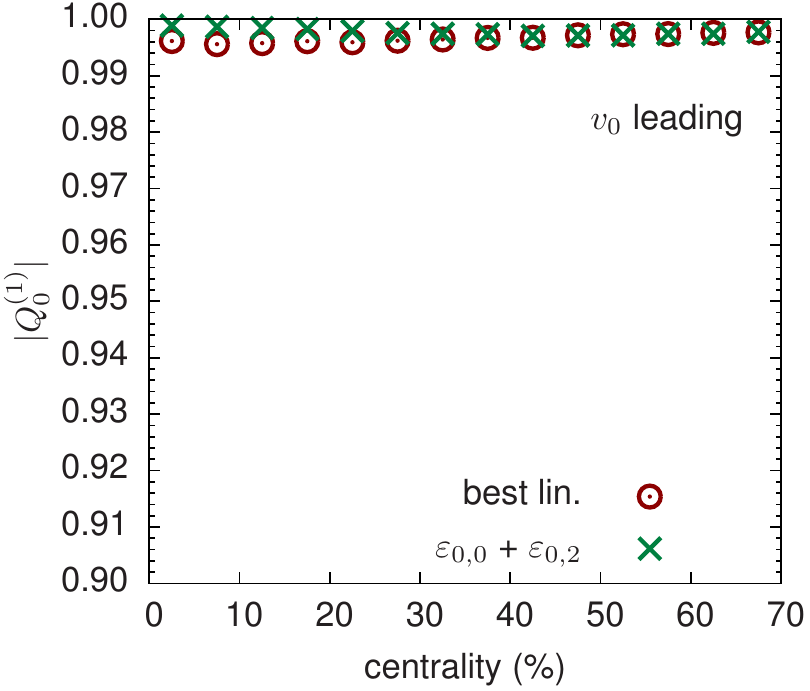}}
\subfigure[{\,Pearson correlation coefficient between the subleading 
flow and several predictors.}\label{v\FLOWN-corr2}]{
\includegraphics{std_A0_a0_B0_b0_corr2_c}}
\caption{$n=\FLOWN$}
\label{v\FLOWN}
\end{figure*}

\renewcommand{\FLOWN}{1}
\begin{figure*}
\centering
\subfigure[
{\,Centrality dependence of the (scaled) magnitudes of flows  
$\|v^{(a)}_\FLOWN\|$.}\label{v\FLOWN-eval}]{
\includegraphics{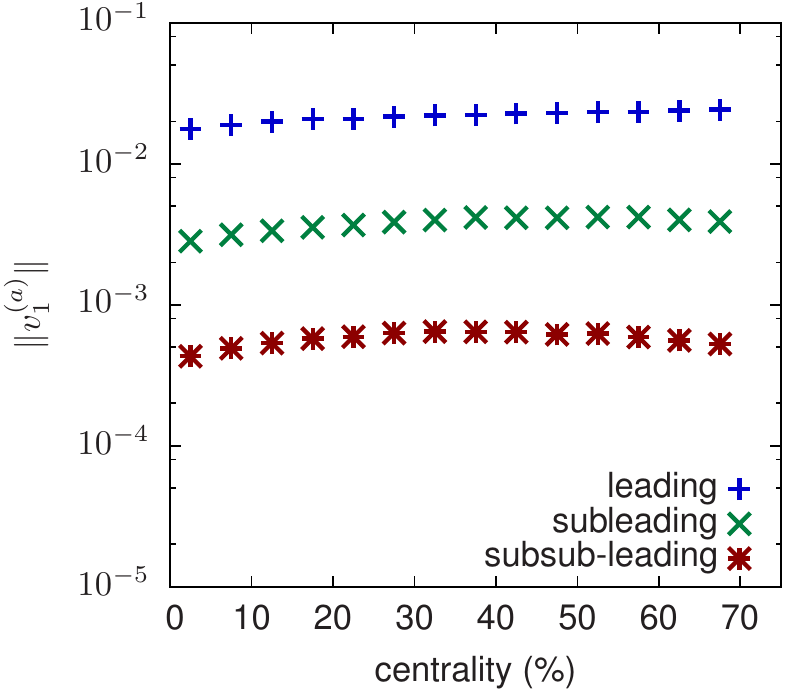}}\\
\subfigure[{\,Momentum dependence of principal flow vectors 
$V_\FLOWN^{(a)}(p_T)$ 
in 
central collisions.}\label{v\FLOWN-Vn_nonorm}]{
\includegraphics{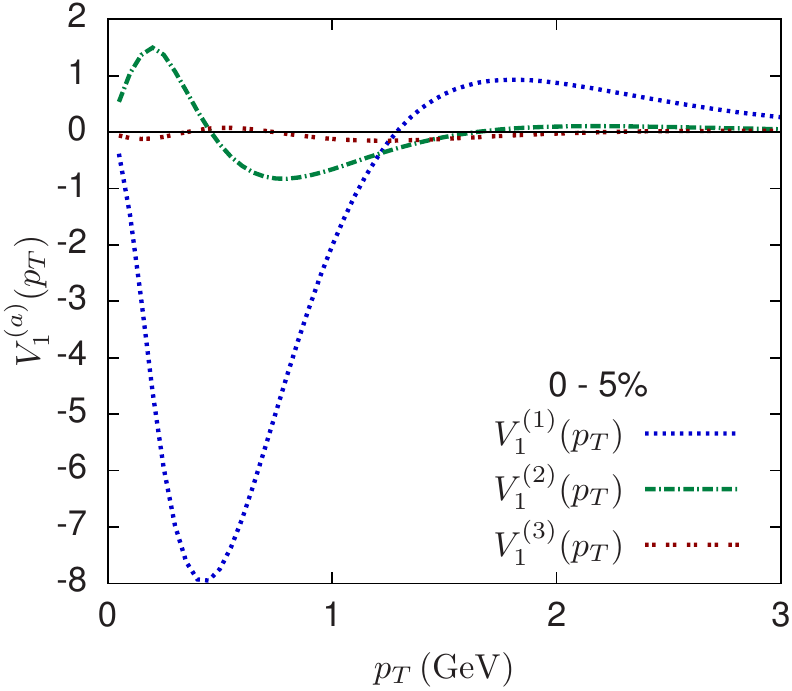}}
\subfigure[{\,Principal flow vectors divided by the average multiplicity, 
$v_\FLOWN^{(a)}(p_T)\equiv 
V_\FLOWN^{(a)}(p_T)/\left<dN/dp_T\right>$.}\label{v\FLOWN-vn_norm}]{
\includegraphics{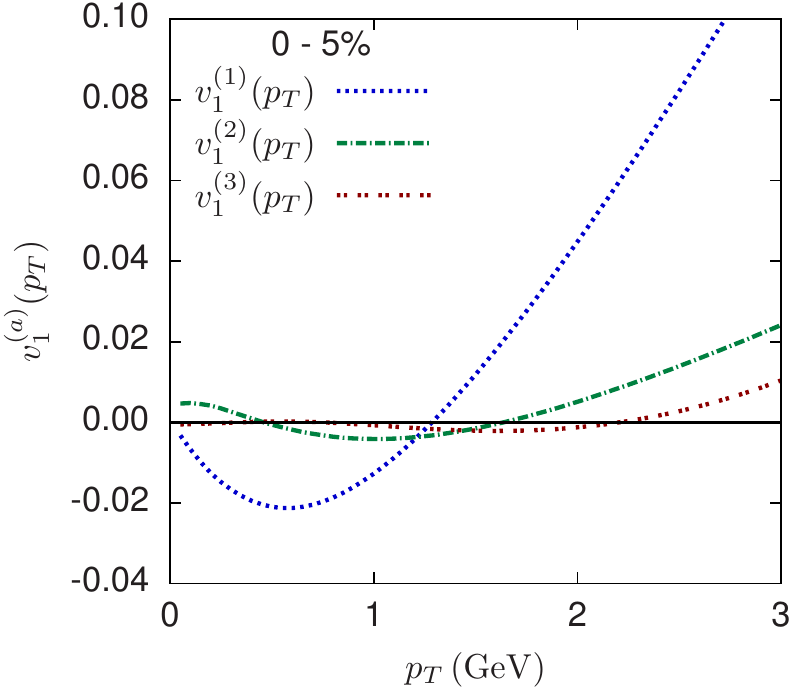}}
\subfigure[{\,Pearson correlation coefficient between the leading flow (zero 
suppressed for clarity) and several predictors.}\label{v\FLOWN-corr1}]{
\includegraphics{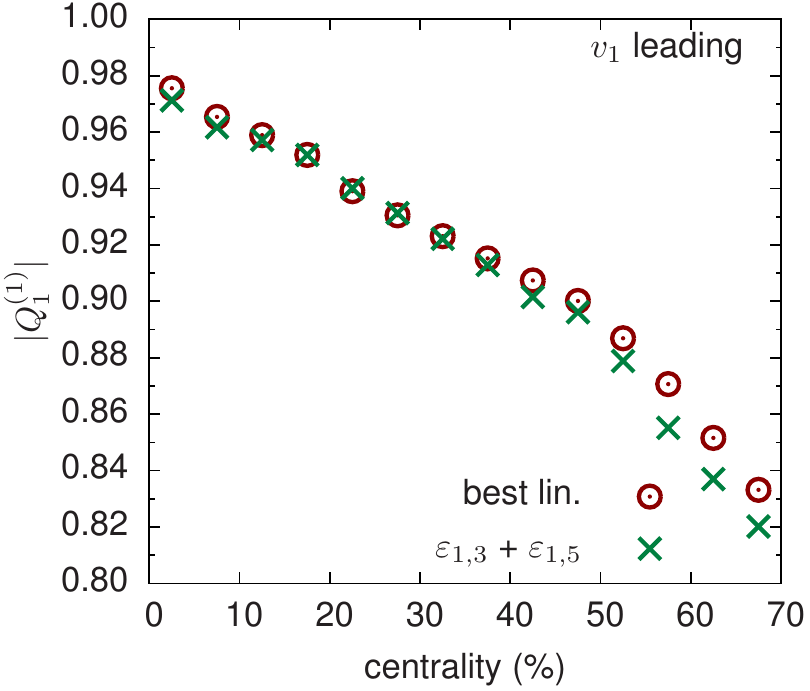}}
\subfigure[{\,Pearson correlation coefficient between the subleading 
flow and several predictors.}\label{v\FLOWN-corr2}]{
\includegraphics{std_A1_a0_B1_b0_corr2_c}}
\caption{$n=\FLOWN$}
\label{v\FLOWN}
\end{figure*}

\renewcommand{\FLOWN}{2}
\begin{figure*}
\centering
\subfigure[
{\,Centrality dependence of the (scaled) magnitudes of flows  
$\|v^{(a)}_\FLOWN\|$.}\label{v\FLOWN-eval}]{
\includegraphics{std_A2_a0_B2_b0_eval_s}}\\
\subfigure[{\,Momentum dependence of principal flow vectors 
$V_\FLOWN^{(a)}(p_T)$ 
in 
central collisions.}\label{v\FLOWN-Vn_nonorm}]{
\includegraphics{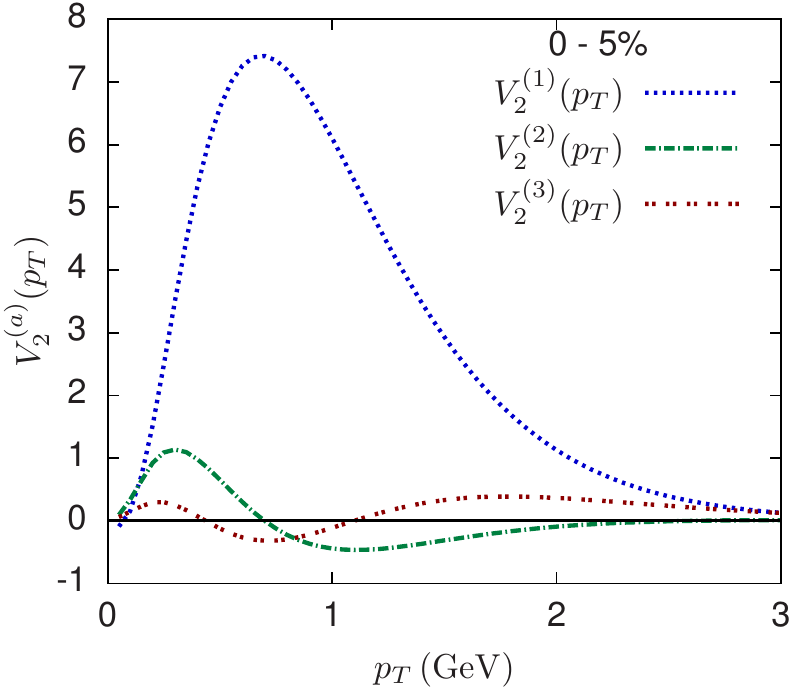}}
\subfigure[{\,Principal flow vectors divided by the average multiplicity, 
$v_\FLOWN^{(a)}(p_T)\equiv 
V_\FLOWN^{(a)}(p_T)/\left<dN/dp_T\right>$.}\label{v\FLOWN-vn_norm}]{
\includegraphics{std_A2_a0_B2_b0_CENT00_vn_norm}}
\subfigure[{\,Pearson correlation coefficient between the leading flow (zero 
suppressed for clarity) and several predictors.}\label{v\FLOWN-corr1}]{
\includegraphics{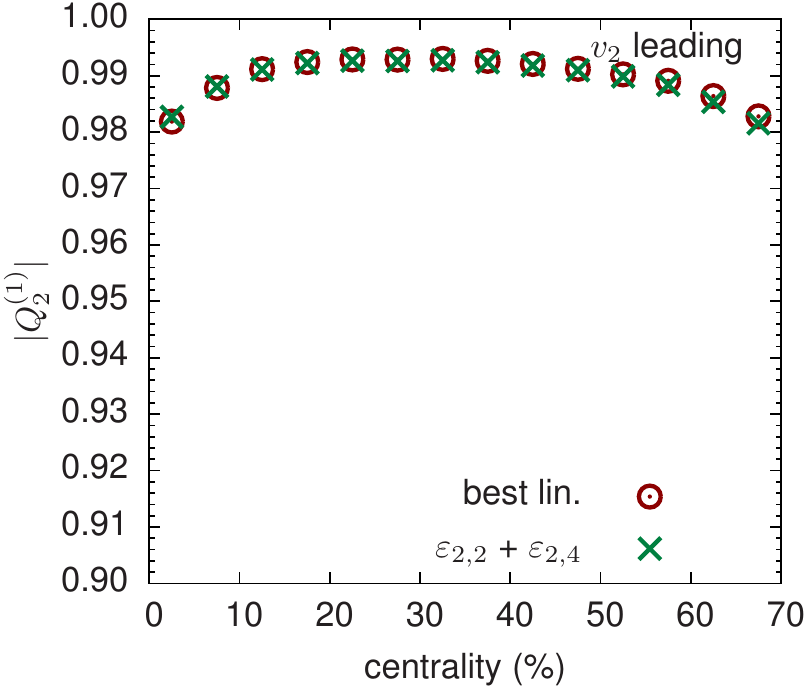}}
\subfigure[{\,Pearson correlation coefficient between the subleading 
flow and several predictors.}\label{v\FLOWN-corr2}]{
\includegraphics{std_A2_a0_B2_b0_corr2_c}}
\caption{$n=\FLOWN$}
\label{v\FLOWN}
\end{figure*}

\renewcommand{\FLOWN}{3}
\begin{figure*}
\centering
\subfigure[
{\,Centrality dependence of the (scaled) magnitudes of flows  
$\|v^{(a)}_\FLOWN\|$.}\label{v\FLOWN-eval}]{
\includegraphics{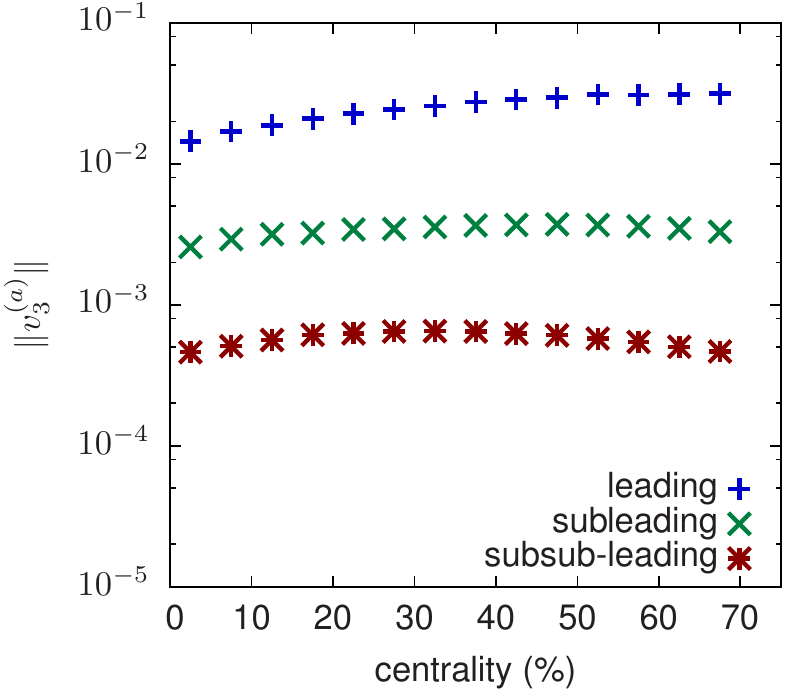}}\\
\subfigure[{\,Momentum dependence of principal flow vectors 
$V_\FLOWN^{(a)}(p_T)$ 
in 
central collisions.}\label{v\FLOWN-Vn_nonorm}]{
\includegraphics{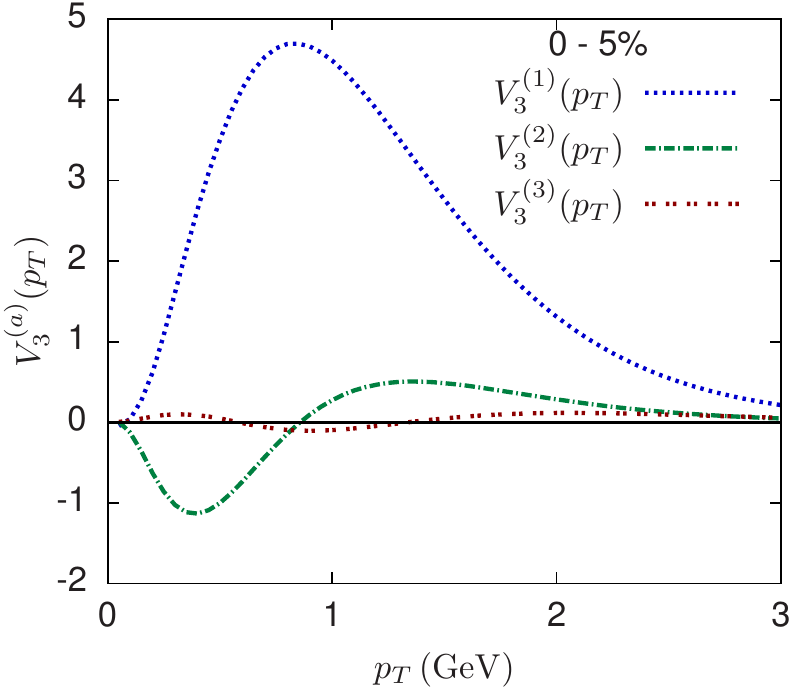}}
\subfigure[{\,Principal flow vectors divided by the average multiplicity, 
$v_\FLOWN^{(a)}(p_T)\equiv 
V_\FLOWN^{(a)}(p_T)/\left<dN/dp_T\right>$.}\label{v\FLOWN-vn_norm}]{
\includegraphics{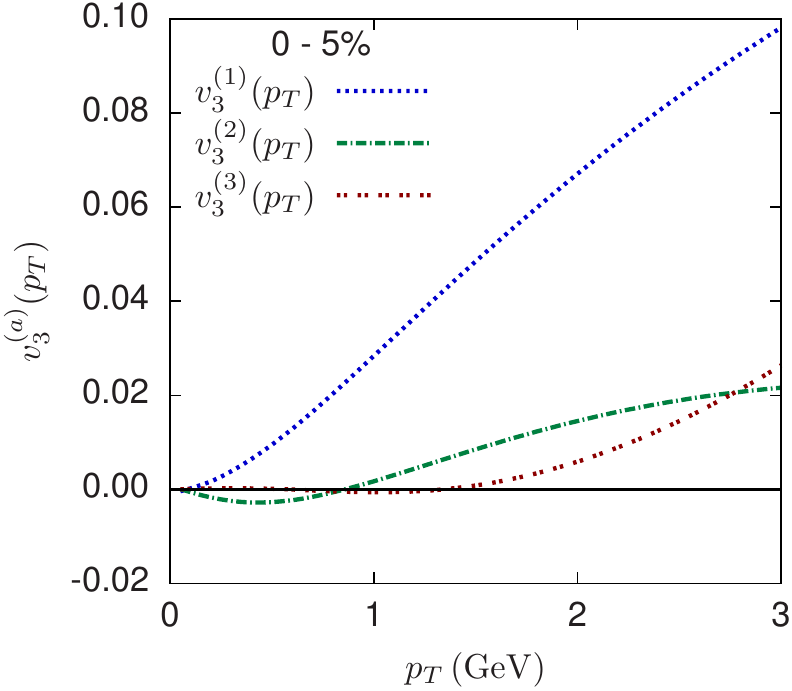}}
\subfigure[{\,Pearson correlation coefficient between the leading flow (zero 
suppressed for clarity) and several predictors.}\label{v\FLOWN-corr1}]{
\includegraphics{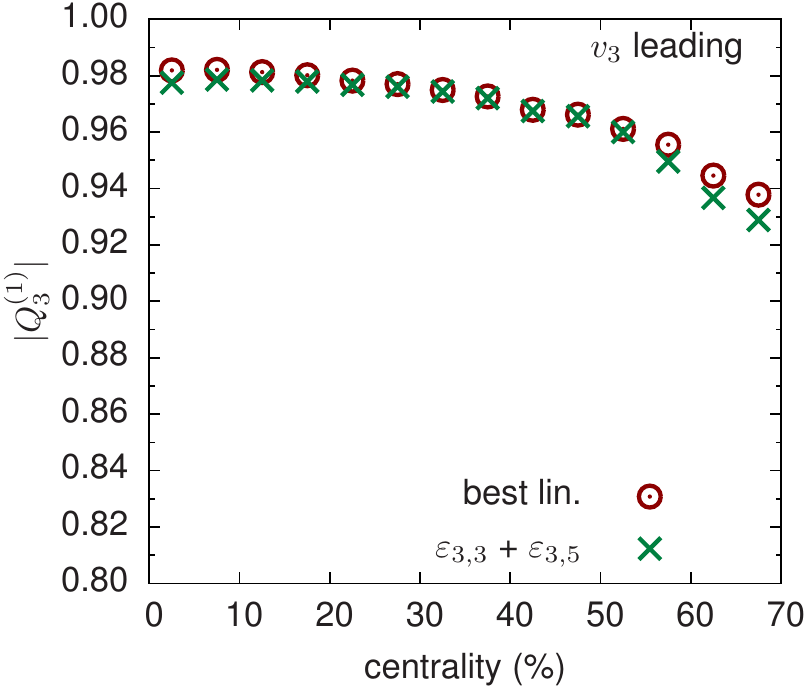}}
\subfigure[{\,Pearson correlation coefficient between the subleading 
flow and several predictors.}\label{v\FLOWN-corr2}]{
\includegraphics{std_A3_a0_B3_b0_corr2_c}}
\caption{$n=\FLOWN$}
\label{v\FLOWN}
\end{figure*}

\renewcommand{\FLOWN}{4}
\begin{figure*}
\centering
\subfigure[
{\,Centrality dependence of the (scaled) magnitudes of flows  
$\|v^{(a)}_\FLOWN\|$.}\label{v\FLOWN-eval}]{
\includegraphics{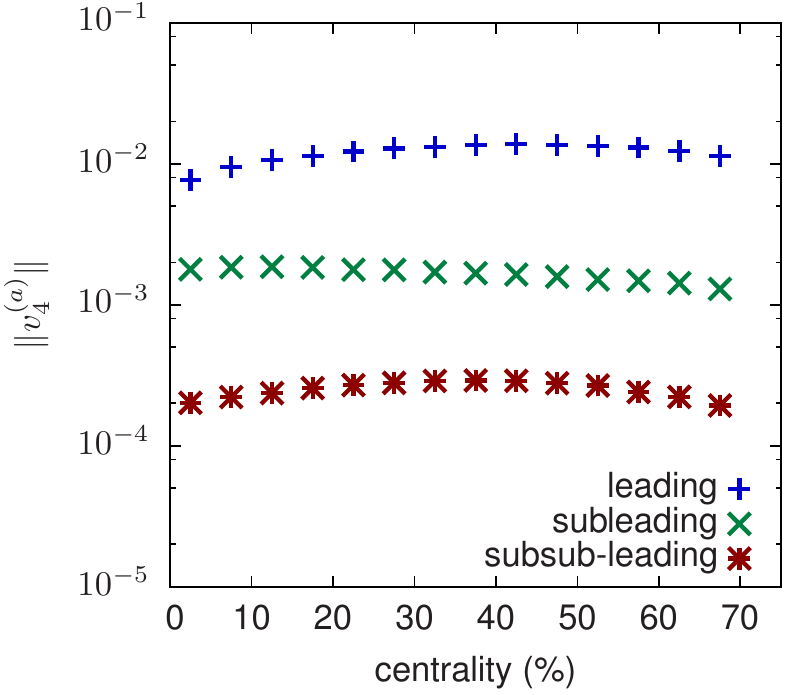}}\\
\subfigure[{\,Momentum dependence of principal flow vectors 
$V_\FLOWN^{(a)}(p_T)$ 
in 
central collisions.}\label{v\FLOWN-Vn_nonorm}]{
\includegraphics{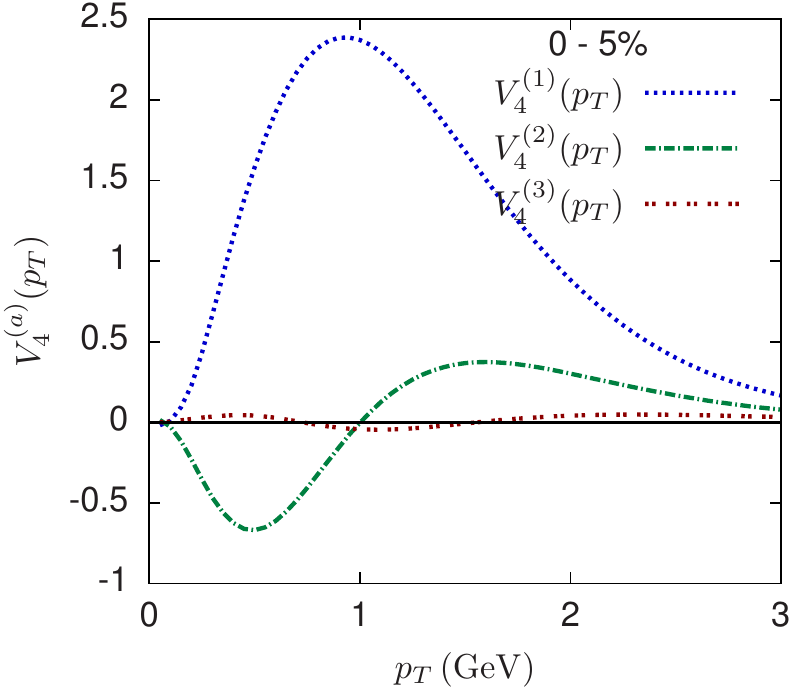}}
\subfigure[{\,Principal flow vectors divided by the average multiplicity, 
$v_\FLOWN^{(a)}(p_T)\equiv 
V_\FLOWN^{(a)}(p_T)/\left<dN/dp_T\right>$.}\label{v\FLOWN-vn_norm}]{
\includegraphics{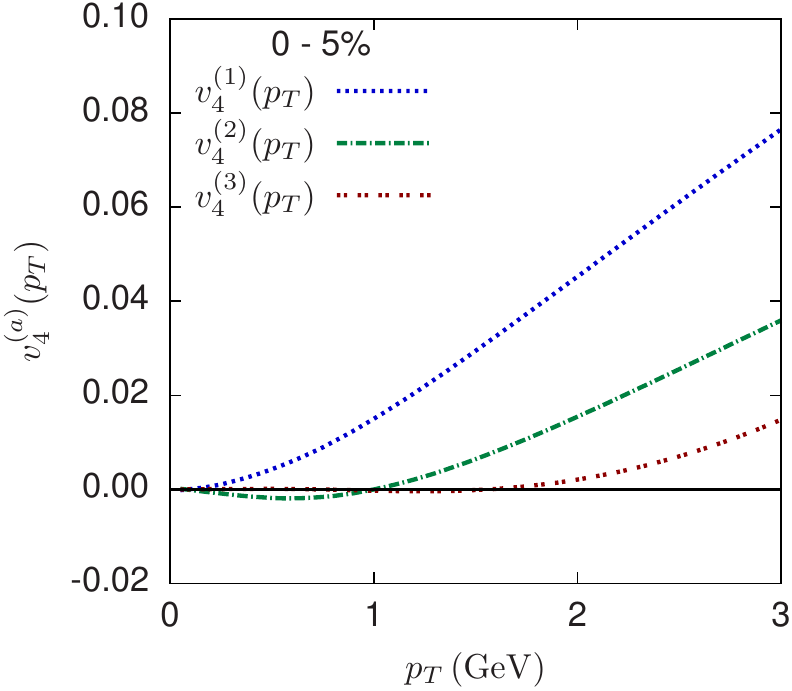}}
\subfigure[{\,Pearson correlation coefficient between the leading flow (zero 
suppressed for clarity) and several predictors.}\label{v\FLOWN-corr1}]{
\includegraphics{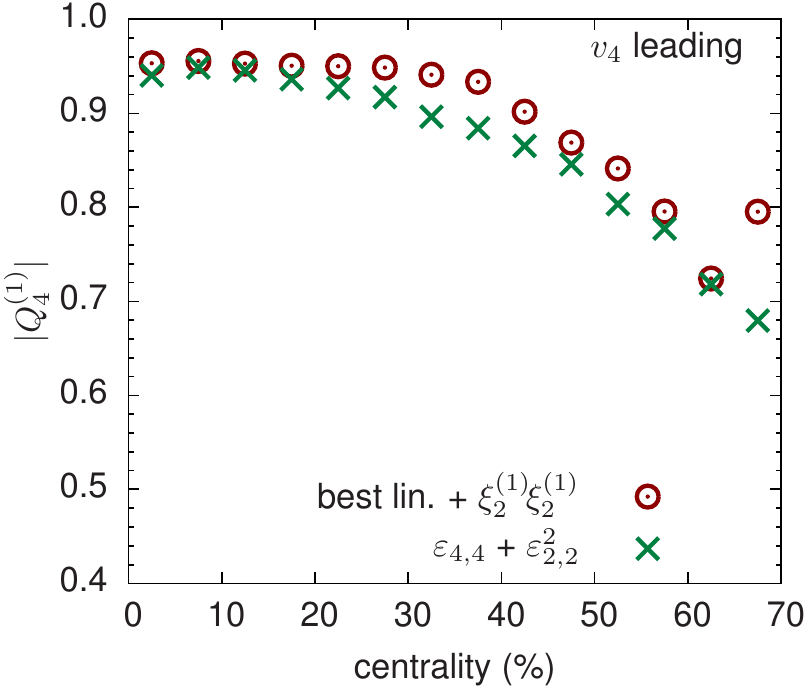}}
\subfigure[{\,Pearson correlation coefficient between the subleading 
flow and several predictors.}\label{v\FLOWN-corr2}]{
\includegraphics{std_A4_a0_B4_b0_corr2_c}}
\caption{$n=\FLOWN$}
\label{v\FLOWN}
\end{figure*}

\renewcommand{\FLOWN}{5}
\begin{figure*}
\centering
\subfigure[
{\,Centrality dependence of the (scaled) magnitudes of flows  
$\|v^{(a)}_\FLOWN\|$.}\label{v\FLOWN-eval}]{
\includegraphics{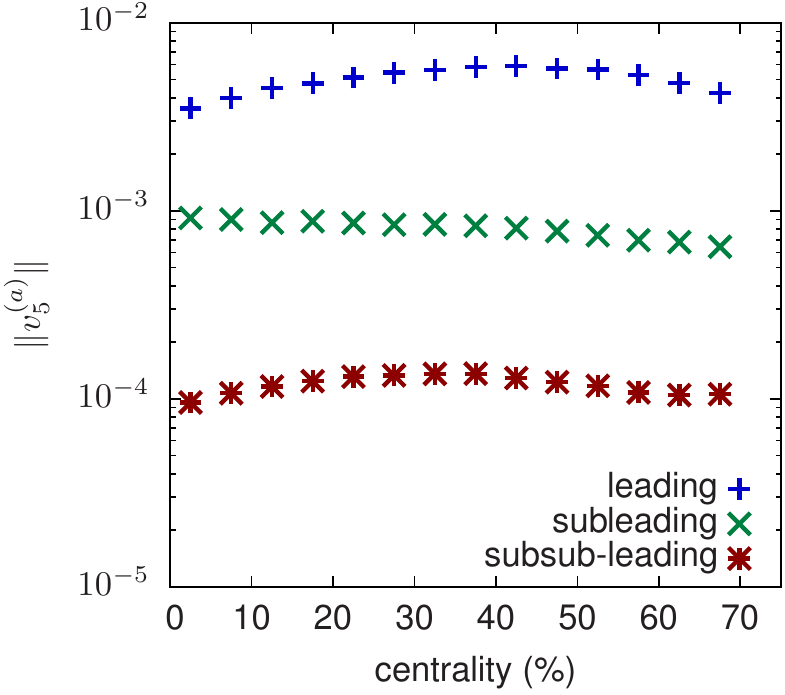}}\\
\subfigure[{\,Momentum dependence of principal flow vectors 
$V_\FLOWN^{(a)}(p_T)$ 
in 
central collisions.}\label{v\FLOWN-Vn_nonorm}]{
\includegraphics{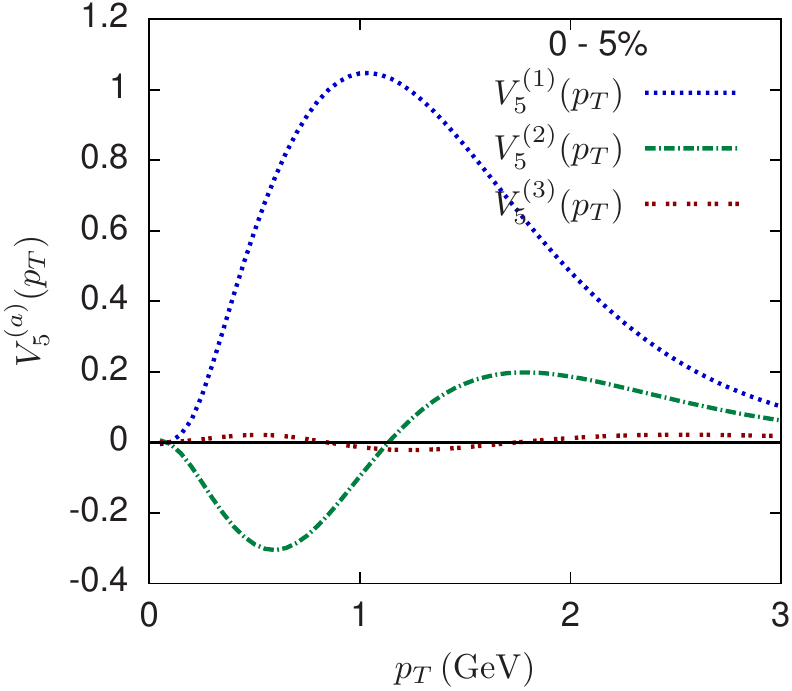}}
\subfigure[{\,Principal flow vectors divided by the average multiplicity, 
$v_\FLOWN^{(a)}(p_T)\equiv 
V_\FLOWN^{(a)}(p_T)/\left<dN/dp_T\right>$.}\label{v\FLOWN-vn_norm}]{
\includegraphics{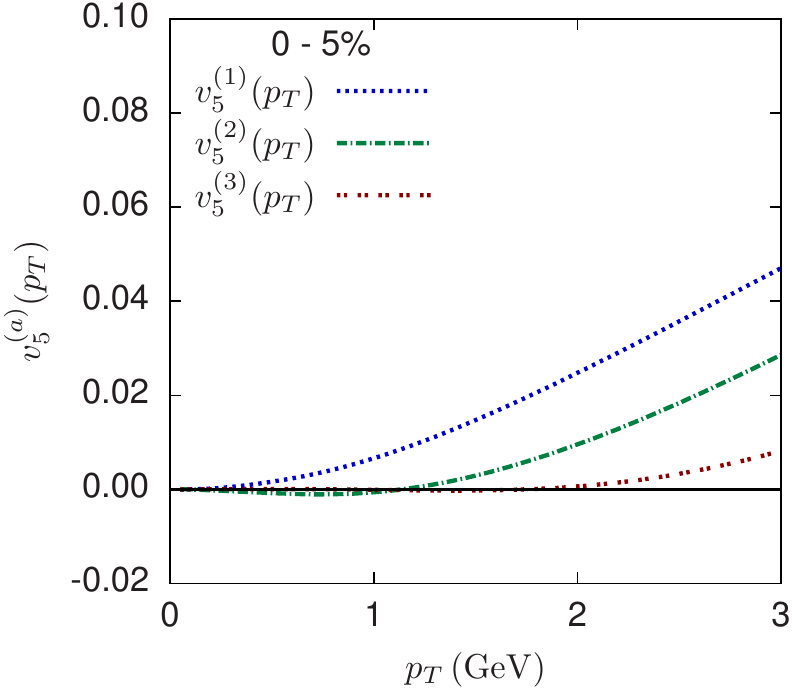}}
\subfigure[{\,Pearson correlation coefficient between the leading flow (zero 
suppressed for clarity) and several predictors.}\label{v\FLOWN-corr1}]{
\includegraphics{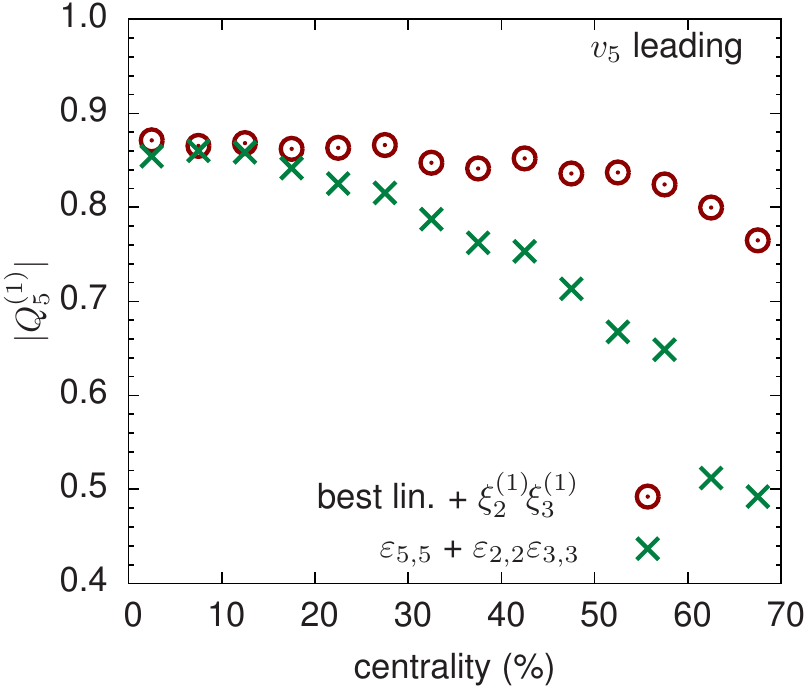}}
\subfigure[{\,Pearson correlation coefficient between the subleading 
flow and several predictors.}\label{v\FLOWN-corr2}]{
\includegraphics{std_A5_a0_B5_b0_corr2_c}}
\caption{$n=\FLOWN$}
\label{v\FLOWN}
\end{figure*}

\end{document}